\documentclass[journal,hideappendix]{vgtc}        

\onlineid{0}

\vgtccategory{Research}

\vgtcpapertype{application}

\title{RCInvestigator: Towards Better Investigation of \\ Anomaly Root Causes in Cloud Computing Systems}

\author{%
Shuhan Liu, Yunfan Zhou, Lu Ying, Yuan Tian, Jue Zhang, Shandan Zhou, Weiwei Cui, \\
Qingwei Lin, Thomas Moscibroda, Haidong Zhang, Di Weng, Yingcai Wu
}

\authorfooter{
  \item
  	Shuhan Liu, Yunfan Zhou, Lu Ying, Yuan Tian, Di Weng, Yingcai Wu are with Zhejiang University.
  \item
    Jue Zhang, Qingwei Lin are with Microsoft Research.
  \item Shandan Zhou, Weiwei Cui, Thomas Moscibroda, Haidong Zhang are with Microsoft.
}

\abstract{%
  Finding the root causes of anomalies in cloud computing systems quickly is crucial to ensure availability and efficiency since accurate root causes can guide engineers to take appropriate actions to address the anomalies and maintain customer satisfaction. However, it is difficult to investigate and identify the root causes based on large-scale and high-dimension monitoring data collected from complex cloud computing environments. Due to the inherently dynamic characteristics of cloud computing systems, the existing approaches in practice largely rely on manual analyses for flexibility and reliability, but massive unpredictable factors and high data complexity make the process time-consuming. Despite recent advances in automated detection and investigation approaches, the speed and quality of root cause analyses remain limited by the lack of expert involvement in these approaches. The limitations found in the current solutions motivate us to propose a visual analytics approach that facilitates the interactive investigation of the anomaly root causes in cloud computing systems. We identified three challenges, namely, a) modeling databases for the root cause investigation, b) inferring root causes from large-scale time series, and c) building comprehensible investigation results. In collaboration with domain experts, we addressed these challenges with RCInvestigator, a novel visual analytics system that establishes a tight collaboration between human and machine and assists experts in investigating the root causes of cloud computing system anomalies. We evaluated the effectiveness of RCInvestigator through two use cases based on real-world data and received positive feedback from experts.
}


\teaser{
  \centering
  \includegraphics[width=\linewidth]{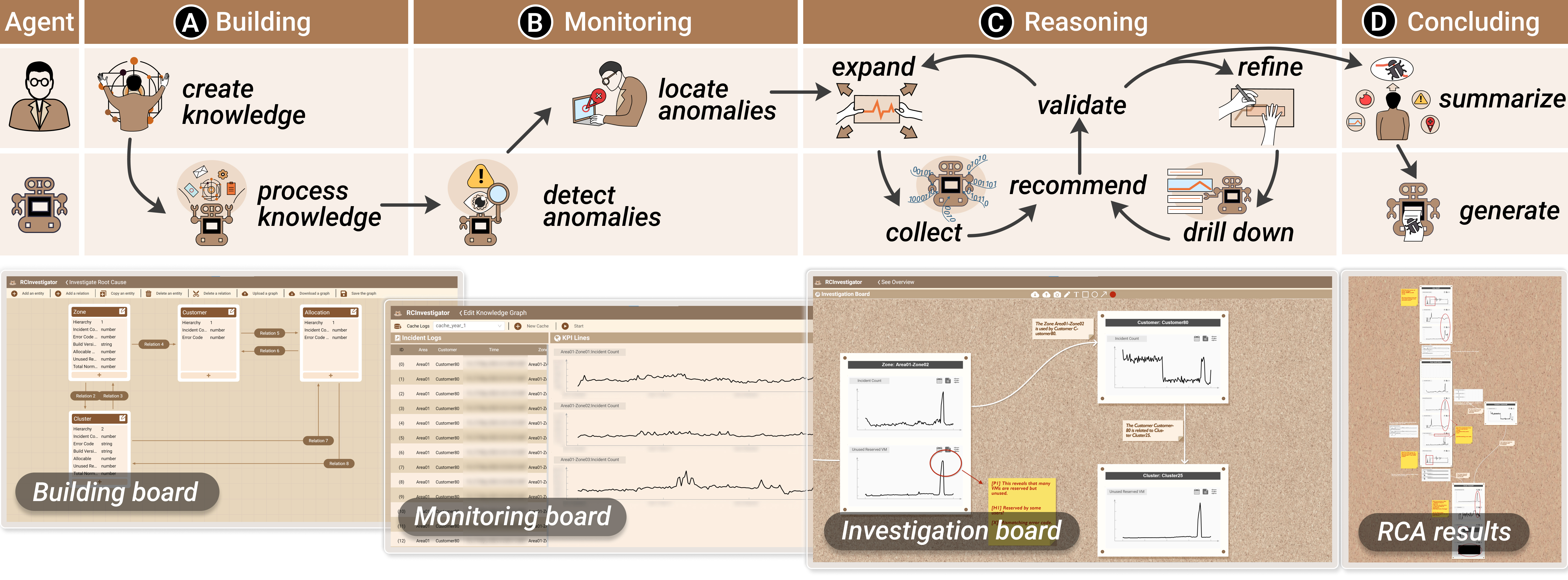}
  \caption{%
    This figure displays a new human-machine-collaborative workflow of root cause analysis in cloud computing systems and an interactive visual analysis system named RCInvestigator.
    The workflow includes four stages: (A) building, (B) monitoring, (C) reasoning, and (D) concluding.
    Two agents (human and machine) collaborate via RCInvestigator, consisting of the building board, the monitoring board, and the investigation board.
  }
  \label{fig:teaser}
}

\graphicspath{{figs/}{figures/}{pictures/}{images/}{./}} 

\usepackage{tabu}                      
\usepackage{booktabs}                  
\usepackage{lipsum}                    
\usepackage{mwe}                       
\usepackage{paralist}
\usepackage{mathptmx}                  
\usepackage{wrapfig}
\usepackage{xurl}
\newcommand{\unsure}[1] {{\color{Black} #1}}
\begin{document}


\firstsection{Introduction}
\maketitle
Cloud computing has emerged as a pervasive infrastructure technology, profoundly influencing a broad spectrum of application domains~\cite{ccbusinesssurvey, hashem2015rise}.
As the significance of cloud computing systems escalates, the prompt detection and management of anomalies become imperative, since these irregularities not only pose potential security risks and performance issues~\cite{IncidentAnomaly} but can also lead to substantial economic losses for companies~\cite{amazonincidents, downtimefor}.
Despite this urgency, the prompt detection of anomalies, pinpointing their origins, and mitigating related risks remain formidable challenges due to the vast volume of monitoring data and the intricate nature of distributed architectures.

While methods for anomaly detection in cloud platforms have advanced considerably~\cite{hagemann2020systematic,MLPDetection,clouddet}, approaches for investigating the underlying root causes of these anomalies are still relatively underdeveloped.
The prevalent root cause analysis (RCA) practice employed in the industry is predominantly dependent on labor-intensive manual processes.
Engineers typically run a series of database queries to gather relevant factors, analyze potential root causes, and deduce the conclusive results. 
Such an analytical procedure is inherently time-intensive, frequently spanning several hours to multiple days to finalize.
Recent studies have attempted to bridge this efficiency gap by developing automated machine learning approaches based on knowledge graphs~\cite{GBRCA,cloudrca} or predefined frameworks~\cite{aadrca}.
Although these methods excel at identifying potentially relevant factors, the diversity of root causes presents a significant challenge in pinpointing the hidden root causes of novel anomalies and offering clear, interpretable explanations that trace the pathway from root causes to anomalies.
Consequently, it is imperative to integrate human expertise into the workflow to enhance the explainability of investigations, provide knowledge-driven summaries, and foster the evolution of automated detection models.

These requirements motivate us to collaborate with experts to develop a solution that facilitates the interactive analyses of anomaly root causes.
Drawing on insights from the existing visual analytics studies on cloud computing~\cite{clouddet,beline}, we propose an interactive approach that synergizes human-computer collaboration.
By harnessing on the nuanced expertise of human analysts and integrating automated techniques for cause factor exploration, the approach aims to improve both the efficiency and the precision of the analytics process.
However, implementing such a solution raises three significant challenges:

\textbf{Modeling complex factor relations for root cause investigation.}
The investigation of anomalies requires factors gathered from multiple databases, 
since the root causes could be dispersed among different platform components.
It is crucial to establishing an accurate data model that reflects the relations among relevant factors to support the consequent analyses.
However, building such a model is naturally challenging due to the scale and complexity of data collected.
Moreover, different analytical intents may require different perspective of the factor relations.
For example, the relation between \textit{users} and \textit{virtual machines} should be \textit{allocation} for capacity anomalies, while such a relation might be \textit{access} for security ones.
To capture such dynamic relations, an intuitive and interactive approach is necessary to incorporate human understanding of complex databases in the modeling process.

\textbf{Inferring obscure root causes from large-scale time series.}
An anomaly typically appears as a spike or drop in the time series recording a key performance indicator (KPI).
To pinpoint the root causes of the anomalies, the engineers will need to analyze and correlate time series quickly to uncover how the KPI is affected by the temporal changes in the root cause factors.
Supporting such investigation is challenging.
On the one hand, the analysts need to identify the factors relevant to the KPI or previously-discovered factors among large-scale monitoring data like finding needles in a haystack;
on the other hand, a comprehensive analysis should be facilitated on the gathered factors to answer questions like how these factors correlate with each other and what conclusions can be drawn.
Therefore, the proposed approach should not only be capable of detecting relevant factors but also providing reasoning support based on massive time series data.

\textbf{Building intuitive presentations of intertwined investigation findings.}
The intuitive presentations of  investigation findings, transcending the current practice of copy-pasting database queries and their results, could facilitate the relevant teams in taking informed and timely actions to address the anomalies.
Furthermore, preserving investigation results as persistent knowledge can support future expansion and inspire further investigation endeavors.
However, transforming the investigative process into an intepretable representation constitutes a formidable challenge.
An investigative process is characterized by its heterogeneity and diversity, encompassing not only exact factors and time series data but also the abstract reasoning logic and expert knowledge that underpin the conclusions drawn.
This complexity inherently complicates the effective dissemination of investigative findings.

In this study, we collaborate closely with the experts from a large cloud service provider and propose an interactive investigation framework based on human-machine collaboration for anomaly root cause analysis.
In response to the aforementioned challenges, our framework comprises four stages, including building, monitoring, reasoning, and concluding, and respectively delegates different tasks to the human and machine agent (see Fig.~\ref{fig:teaser}), whereas the machine agent is responsible for laborious data collection tasks and the human agent is responsible for decision-making.
First, knowledge graphs are employed to model the complex factor relationships across multiple data sources.
Then, we design intuitive visualizations and a layout algorithm to organize the factors reasonably for visual analysis.
Finally, we summarize two types of reasoning interactions and design corresponding reasoning models to enable steerable reasoning in finding the root causes of anomalies.

Our main contributions are listed as follows.
\begin{compactitem}
\item We characterize the user requirements for analyzing anomaly root causes in cloud computing systems and formulate a human-machine-collaborative root cause investigation framework.

\item We propose a novel knowledge-graph-based interactive approach for root cause analysis and develop RCInvestigator, an interactive system that supports steerable reasoning of the root causes.

\item We evaluate RCInvestigator with two use cases based on a real-world dataset and collect qualitative feedback from experts.
\end{compactitem}
\section{Related Work}
We survey related studies from three perspectives: root cause analysis, time-oriented data visualization, and complex system visual diagnosis. 

\subsection{Root Cause Analysis}
Since root cause analysis has been regarded as one of the most challenging tasks in system operations, many researchers reviewed RCA approaches from different angles, such as general aspects (e.g., ~\cite{surveyRCAGeneral},~\cite{surveyRCAInfer}) and specific domains (e.g., software~\cite{surveyRCAsoftware}, industry~\cite{gao2015survey}).
Due to the distributed native of cloud computing, we mainly concentrate on methods for solving RCA of data anomalies in cloud environments. 
We followed the survey~\cite{cloudRCAsurvey,cloudrca}, dividing existing approaches into three folds: manual-based, correlation-analysis-based, and graph-based.

\textbf{Manual-based} methods usually identify root causes by analyzing log and KPI data from various levels of applications, operating systems, networks, and infrastructure.
Although automated log tracking methods (e.g., ~\cite{aadrca,alog}) can quickly locate beginning anomalous logs, providing insightful root-cause explanations is still difficult due to the limited types of logs.
Many cloud computing analysts often combine log analysis with further queries of database information to complete the analysis in practice. 
However, manual log analysis is labor-intensive and often takes several hours or even days to locate the root cause.

\textbf{Correlation-analysis-based} methods compute the correlation coefficient between predefined KPI data collected from monitors~\cite{cloudRCAsurvey}, supporting the investigation of application-~\cite{FChain, PAL}, service-~\cite{wangservice,traceanomaly}, and platform-level~\cite{shanplatform, corr} root causes.
However, these methods do not fully leverage the domain knowledge of infrastructures and databases thus limiting their ability to give an interpretable reasoning process.

\textbf{Graph-based} methods construct graphs via unsupervised and supervised learning and find potential root causes through walking on built graphs.
The built graphs include topology graphs~\cite{monitorrank}, causality graphs~\cite{causeinfer2014, causeinfer2019}, and knowledge graphs~\cite{cloudrca}.
The walking strategies can be Breadth-First-Search~\cite{lin2018microscope}, Markov analysis~\cite{DLA}, KPI-correlation-oriented~\cite{Sieve}, and random walk~\cite{MicroRCA}.
Unsupervised learning methods require no additional labels, but their precision and recall rate are limited.
Although supervised learning methods perform well in finding known types of root causes, obtaining labeled data is extremely difficult and costly, especially with large-scale data in cloud environments.
Besides, existing known labels are only the peak of an iceberg compared with unknown root causes, which means human knowledge must be involved in the investigation loop to improve models' evolution ability.

Though existing automated approaches ensure real-time performance, they lack interpretability and controllability. 
Thus, we introduce human intelligence in the investigation loop and achieve effective human-machine collaboration in RCA tasks. 

\subsection{Time-oriented Data Visualization}
Numerous studies discussing temporal data visualizations from different perspectives, like general time-oriented data~\cite{AIGNER2007401, aigner2023visualization}, discrete sequential events~\cite{eventsurvey}, and continuous time series~\cite{Fang_2020}. 
Data collected from cloud environments are mostly linear, including time intervals (events) and time points (states).
Displaying the relations between events and states is important for RCA, so we mainly discuss visualizations for event sequences and time series as well as their relations.

As for visualizing time-oriented data, techniques can be categorized into three types: chart-, timeline-, and tree-based.
Chart-based methods (e.g., RetainVIS~\cite{RetainVis}) visualize the distribution of event or point instances by basic bar charts or scatter plots. 
Timeline-based techniques (e.g., PlanningVis~\cite{planningvis}) present the order of events and values consistently alone a timeline and emphasize the temporal features.
Tree-based methods usually display the aggregation patterns (e.g., VisRuption~\cite{rosenthal2013visruption}) or the hierarchical patterns (e.g., VizTree~\cite{lin2004viztree}).

As for visualizing the relations and evolutions, there are general techniques and special designs.
General high-dimensional data visualizations (e.g., PCP~\cite{tominski2004axes}) can display relations between different attributes but may fall short in describing temporal features.
Special-designed visualizations (e.g, TimeCurve~\cite{timecurve} and Sankey ~\cite{wongsuphasawat2011outflow}) intuitively reveal the evolution but are for events or time series specifically.

Based on extensive literature research, RCInvestigator adopts the most basic and easy-to-understand visualizations: Gantt charts for event sequences and line charts for time series. 
We further design novel visualizations and frameworks that intuitively highlight the relations among event sequences and time series, supporting RCA tasks.

\subsection{Complex System Visual Diagnosis}
Complex system visual diagnosis integrates intuitive visual representations, flexible interactions, and effective models to enable a comprehensive understanding of anomalous patterns and large-scale data.
It is effective in solving problems in complex systems, covering various domains, such as industry~\cite{manu2019survey, industrysurvey}, urban computing~\cite{deng_survey_2023}, and dynamic network analysis~\cite{networksurvey, patient, citenetwork}.
We discuss aspects of pipeline risk management through the lens of a general workflow~\cite{riskpipeline}, specifically addressing monitoring, anomaly detection, and root cause analysis.

In the stage of \textbf{monitoring}, the main challenge of visualizing such data is to organize large-scale and high-dimension data reasonably.
The commonly used visualization techniques include static and dynamic methods.
Many static approaches not only utilize dimension-reduction models and cluster algorithms~\cite{Hardware} but also adopt high-dimension visual representations, such as small multiples~\cite{beline}.
Dynamic approaches introduce interactions (e.g., Traveler~\cite{Traveler}) and animations (fish eyes, dynamic PCP) in visualizations and enable level-of-detail displaying, making the monitoring process more flexible than static ones.

In the stage of \textbf{anomaly detection}, visual analysis has been proven as an effective and accurate method in many fields. 
In the industrial sector, ViDX~\cite{ViDX} utilizes Marey charts to highlight anomalies, while ECoalVis~\cite{ECoalVis} enables interactive extraction of anomalies with trend events.
In software analysis, CloudDet~\cite{clouddet} uses clustering algorithms and glyph-based visualization to show the multidimensional features of anomalies and supports interactive detection, while GRANO~\cite{GRANO} locates anomalous components through a knowledge graph. 
ViSRE~\cite{ViSRE} leverages causal inference to predict anomalies proactively.

These techniques showcase the power of visual analysis in monitoring and detecting anomalies across various domains.
However, few studies have focused on utilizing interactive visual analysis methods to achieve root cause analysis of complex system anomalies. 
Therefore, we propose the RCInvestigator framework and develop a system.
\section{Background and Task Abstraction}
\subsection{Background}
In our study, the development of RCInvestigator involved an iterative design process that was guided by Sedlmair et al.'s design study methodology~\cite{designmethod}.
Such a design process unfolded over a six-month period and involved three steps: (1) first, we conducted a comprehensive review of existing commercial tools and visual analytics systems to understand the current research landscape;
(2) over the following two months, we engaged in multiple interview sessions with four domain experts (EA-D) to obtain their nuanced insights and summarized pain the points and user requirements in the RCA processes;
(3) in the final three months, we embarked on an iterative design process, where we created preliminary visual design mockups and refined them over three iterative rounds in collaboration with the experts, ensuring that the final system design was both user-centric and aligned with expert feedback.
All of the experts were employed by a leading cloud computing provider.
EA (female) and EB (male) were data scientists with years of experience in conducting cloud platform incident analyses, while EC (male) and ED (male) were senior researchers specialized in AIOps, facilitating IT operations in cloud computing systems with AI-based approaches.

\begin{figure}[b]
  \includegraphics[width=\linewidth]{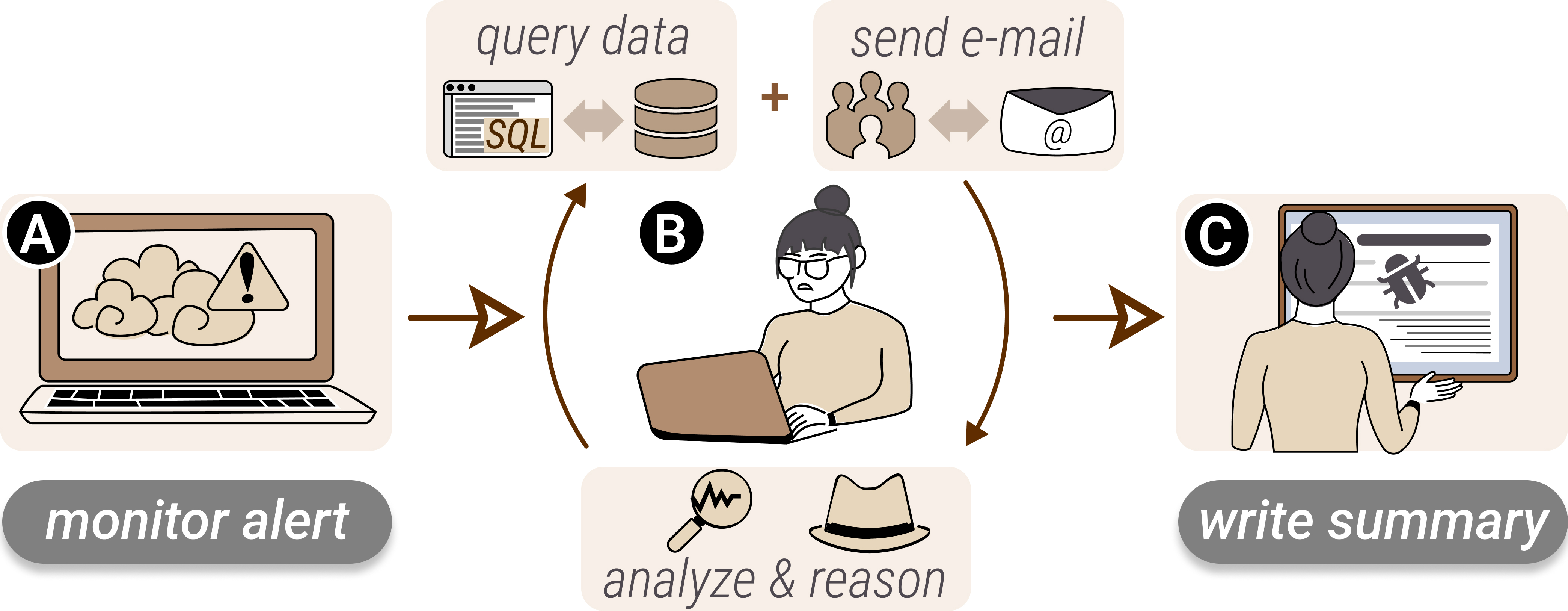}

  \caption{The existing RCA pipeline includes three stages, mainly relying on manual efforts. 
  (A) First, analysts monitor the list of recent alerts and decide which one should be analyzed. 
  (B) Second, analysts write query scripts to retrieve data and e-mail relevant teams for additional information.
  Then, analysts reason possible causes and gain insights. 
  They repeat the process until the root cause is identified.
  (C) Finally, analysts write a summary.
  }
  \label{fig:old_pipeline}
\end{figure}

During months of collaboration with the experts, we have gained insights into the process of investigating anomaly root causes in cloud computing systems. 
The existing workflow, as shown in~\cref{fig:old_pipeline}, can be divided into three stages: 
First, analysts monitor an incident alert list and decide which incident is urgent to response.
Second, analysts query databases for relevant data based on personal experience and send e-mails to relevant teams for assistance. 
After gaining insights by analyzing the collected clues, analysts repeat the querying and analyzing procedure until a possible root cause is identified. 
Finally, analysts write an RCA summary, which is important for sharing insights on the incident and potential mitigation actions to responsible teams and other colleagues and also for persisting RCA experience as knowledge to accelerate future investigations of similar incidents.
Through a series of interviews, three pain points were identified in such a workflow.

\begin{compactitem}
\item [\textbf{P1.}] The collection process of potential clues is cumbersome.
Analysts need to write query scripts manually to retrieve relevant data from databases, but it is hard to have a comprehensive understanding of such complex and large-scale databases.
Furthermore, there is much repetitive work like writing numerous similar and lengthy scripts, making the process time-consuming and labor-intensive.
\item [\textbf{P2.}] The reasoning process causes a heavy mental burden.
Analysts must delve into the collected clues and compare them one by one to make inferences, which incurs a high memory cost.
Moreover, existing representations of clues are mainly raw tabular data, which is not intuitive, causing additional cognitive difficulty.
\item [\textbf{P3.}] The investigation results are hard to share.
Current analysis summaries are mainly composed of scattered information like text and query results, which fails to present the analysts' reasoning logic in an organized and intuitive manner.
Additionally, the arbitrary format of the results makes it difficult to share the knowledge.
\end{compactitem}

To solve these pain points, we propose a new human-machine-collaborative RCA workflow.
We delegate repetitive collection tasks to the machine agent (\textbf{P1}), empower the reasoning by designing an interactive visual approach to bridge human and machine agents (\textbf{P2}), and build all these on a structured investigation consensus to facilitate sharing (\textbf{P3}).
In the new workflow (see~\cref{fig:teaser}), there are one preparation stage (\textit{building}), three investigation stages (\textit{monitoring}, \textit{reasoning}, and \textit{\unsure{concluding}}) and two agents (\textit{human} and \textit{machine}).
In each stage, two agents perform different tasks:
\begin{compactitem}
    \item In the \underline{Building} stage, a consensus, which describes how human and machine agents collaborate, should be defined.
    Specifically, the consensus is the shared knowledge of databases, anomalies, root causes between human and machine agents. 
    Users dominate the definition of consensus to ensure its interpretability and reliability, while the machine agent follows it to adapt data loading.
    \item In the \underline{Monitoring} stage, automated models in the machine agent \textit{detect} potential anomalies and alarm users, and then users \textit{locate} a detailed anomaly, which they want to investigate its root cause.
    \item In the \underline{Reasoning} stage, users first \textit{propose} hypotheses based on the selected anomaly, and then the machine \textit{collects} as many clues as possible from massive data and \textit{recommends} them based on the degree of association. 
    Users then \textit{validate} recommended clues to identify which are true evidence.
    With new pieces of evidence, users may be inspired and propose new hypotheses, and the machine repeatedly collects clues.
    The second stage is iterative until identifying final root causes.
    There are two kinds of hypotheses, one is for expanding the scope of clues, and the other is for refining clues.
    When users \textit{expand} a clue, the machine \textit{searches for} relevant clues as many as possible to generalize the analysis.
    When users \textit{refine} a clue, the machine \textit{drills down} the clue and eliminates unnecessary factors to detail the investigation. 
    
    \item In the \underline{Concluding} stage, users \textit{summarize} final root causes via annotations, and the machine \textit{generates} structured and visual format of conclusions for share.
\end{compactitem}

\subsection{Data and Concepts}
\label{sec:data}
This study is based on the data obtained from the cloud service provider where the experts were employed.
Several terminologies are introduced.

In the building stage, two types of anomalies are involved: (1) \textit{Incident logs} record a log that a user failed to request virtual resources. Each incident log is a tuple of number and string; (2) \textit{Key performance indicators} record states of components and are numerical time series. 

In the reasoning stage, clues can be classified into two categories: (1) Time series, which record numerical changes in components and performance metrics; (2) Event sequences, which record categorical changes in options.
Validated clues are regarded as pieces of evidence.
Data types include \textit{number} (a time series), \textit{string} (an event sequence), \textit{set} (multiple event sequences), and \textit{bag} (multiple time series).

\subsection{Requirements Analysis}
\label{sec:req}
We summarize five user requirements as follows.
Each user requirement is a refinement of the investigation stages in the new workflow. 
Specifically, \textbf{R1} corresponds to the building stage, \textbf{R2} to the monitoring stage, \textbf{R3} to the reasoning stage (the human agent), \textbf{R4} to the reasoning stage (the machine agent), and \textbf{R5} to the concluding stage.

\textbf{R1. Build persistent investigation knowledge interactively.}
RCA relies on investigative knowledge, which requires both human and machine agents to understand \textit{which clues can be causes} and \textit{how to collect clues}.
The model of investigative knowledge should effectively distinguish and connect these two types of clues.
Besides, due to the sheer volume of data and the complexity of infrastructure, the proposed system should ensure the accuracy and reliability of the knowledge to reduce the query cost.
Thus, interactive building as well as flexible extension of the knowledge should be considered.

\textbf{R2. Obtain an overview of cloud computing anomalies.}
In the monitoring stage, experts focused on gaining a high-level understanding of cloud computing systems' performance.
They may want to learn about the KPIs in key areas, such as allocable resource volumes.
Therefore, an intuitive overview of KPIs should be provided.
Additionally, the experts are also interested in incident logs, which reflect necessary clues for anomaly cause investigation, so the proposed system should also facilitate inspecting incident logs.
Moreover, the proposed system should display anomalies on demand due to the large scale of data.

\textbf{R3. Extend analysis scope based on recommendations.}
In the reasoning stage, it is crucial for experts to efficiently extend the analysis scope by proposing and validating hypotheses based on recommended clues.
On one hand, the proposed system should present recommendations in an intuitive manner and assist analysts in gaining valuable insights that can inspire further reasoning. 
Additionally, these clues should be organized in a clear and interpretable way, enabling direct comparison. 
On the other hand, it is necessary to integrate new interactions that facilitate hypothesis proposing and validating, thus enabling full leverage of existing investigative knowledge.

\textbf{R4. Explore time-oriented data based on investigation knowledge.}
In the reasoning stage, users require the machine agent to collect as many relevant clues as possible and provide the ones that are most likely to lead to the root cause.
On one hand, the collected clues must be highly interpretable, so the collection process must be closely integrated with investigative knowledge. 
On the other hand, there must be a reasonable way to evaluate and describe the degree of correlation between the two and provide reliable recommendations because of the complexity and heteregenousity of clues.
Furthermore, it is important to ensure that the collection and recommendation processes are real-time to avoid interrupting the user's investigation.

\textbf{R5. Share intuitive investigation results.}
In the concluding stage, the investigation results must be formed to an expressive output that supports free sharing. 
To make the output more expressive and interpretable, it involves designing an interface that allows users to add notes, highlight key findings, and provide summaries on the results. 
Exporting the output with such diverse elements into a visual format that facilitates understanding is also important.
Moreover, for further collaboration and mitigation, the proposed system should structure the output into a standardilized format, which is easy to store and reload. 

\section{RCInvestigator}
To fulfill the user requirements summarized in Sec.~\ref{sec:req}, we followed the four-stage workflow and proposed RCInvestigator, an interactive approach supporting human-machine-collaborative investigation of anomaly root causes in cloud computing systems.
RCInvestigator is a web-based application based on JavaScript-Vue~\cite{vue}, Python-Flask~\cite{flask}, and Kusto~\cite{kusto}. 
RCInvestigator consists of a frontend interface, a backend model, and a data loader.
(a) The frontend interface is composed of three boards, namely, the building board, the monitoring board, and the investigation board.
(b) The backend model supports dealing built knowledge into queries, retrieving detected anomalies, collecting and recommending clues, and generating summaries.
(c) The data store connects to databases and keeps caches.
In this section, we introduce the design and implementation details for each stage.

\subsection{Build Investigation Knowledge}
RCInvestigator adopts the knowledge graph, which can be stored in a structured form and kept persistently as well as flexible enough to expand, as the representation of investigation knowledge (\textbf{R1}).
We will introduce the investigation knowledge graph model and the building board that supports the interactive creation and editing of knowledge.

\subsubsection{Model investigation knowledge graph}
A knowledge graph is typically defined by three components: entities, relations, and attributes.
To model the investigation knowledge into a graph, we discuss how to match different knowledge with each graph component.
There are two types of investigation knowledge: (1) Cause clues: anomalous attributes that indicate possible root causes; (2) Reasoning logic: facts that reveal how to collect cause clues.

\textbf{\textit{Cause clues.}}
After observing a large amount of root cause examples, we found that a cause clue can be represented as an observation on a tuple of (entity, attribute, time range).
In this tuple, the (entity, attribute) reveals \textit{``where''} the clue is, while (time range) reveals \textit{``when''} the clue happens. 
For example, the root cause that \textit{``The nodes of cluster A are used up during 2:00 to 3:00 so anomalies happen''} can be simplified as a tuple like \textit{(cluster A, node count, 2:00-3:00)}.
If the time range is given, the cause clue can be defined by (entity, attribute), which is similar to a component of a knowledge graph.
The entity is an instance in a real cause clue (e.g., Cluster A) but is an abstract concept in the knowledge graph (e.g., Cluster). 
To simplify the narration, \textit{``entity concept''} means the abstract concept but \textit{``entity''} means the instance.
Also, we use \textit{``clue''} for short.
Formally, clues can be presented under a given time range $t$. 
We use $E$ to present the set of entity concepts.
Each entity concept $e_i\in E$ has a set of attributes $A_i$, so a cause clue is $(e_{iq}, a_{ip}, t)$, where $ a_{ip}\in A_i$ and $e_{iq}$ is an instance of $e_i$.
To make attributes specific, users can define filters $B$.
For each filter $b_i\in B$, there is a set of options $C$.
If users do not predefine filters, attributes of the data type string will be regarded as default optional filters.

\textbf{\textit{Reasoning logic.}}
They reveal the relations between clues based on facts.
A typical fact can be a zone contains many clusters so if analysts find anomalies in a zone, they can also collect clues from its clusters.
Reasoning logic can correspond to relations in a knowledge graph.
Formally, the set of reasoning logic is $R$ and the set of facts is $F$.
Each fact $f_k \in F$ is tuple $(e_i, r_k, e_j)$, where $r_k \in R$ connects $e_i \in E$ and $e_j \in E$.
The meaning is $f_k$ indicates $e_i$ has a relation $r_k$ with $e_j$. 
In conclusion, the investigation knowledge graph is $G=(E, R, F)$.

\begin{figure}[t]
  \includegraphics[width=\linewidth]{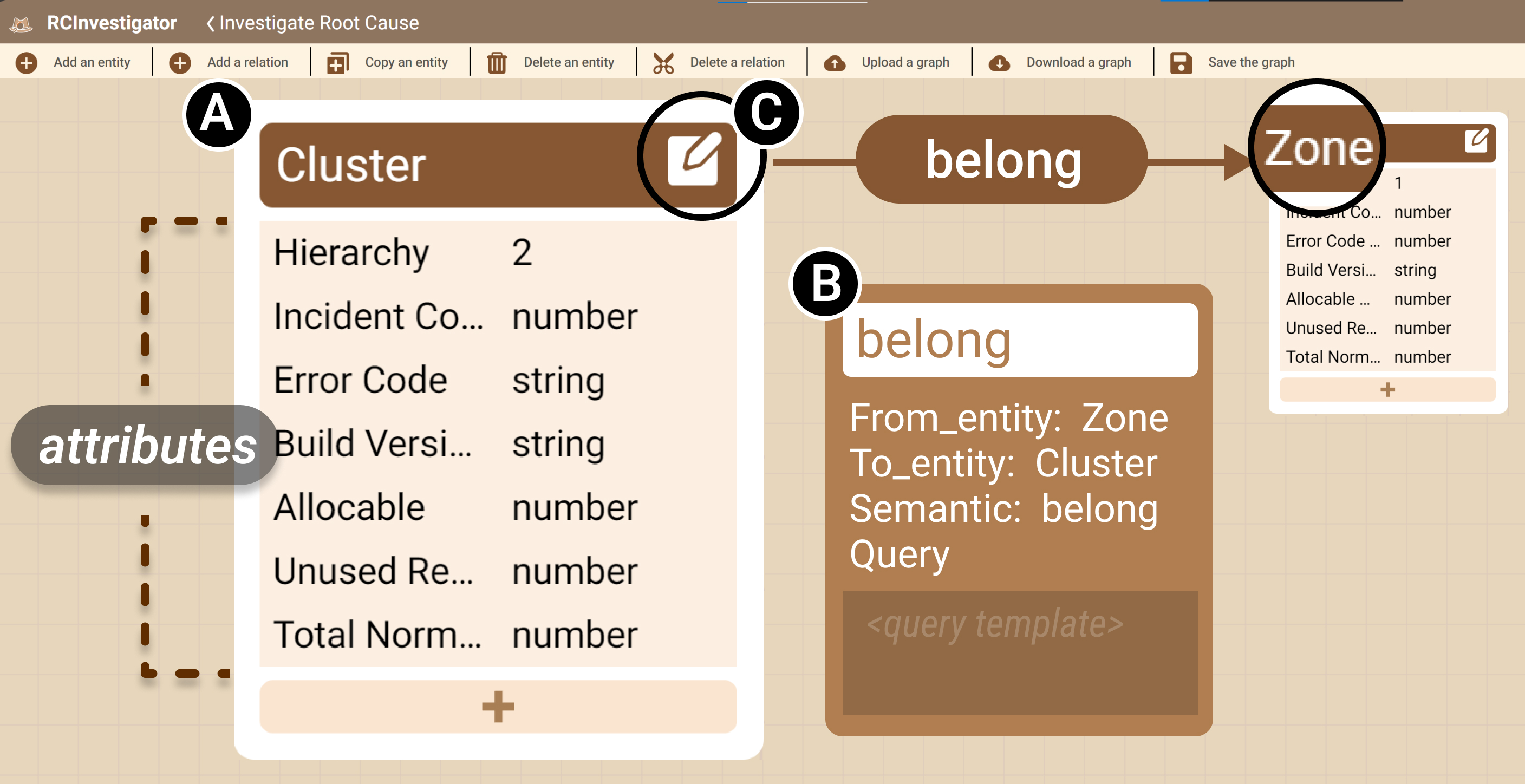}
  \caption{The building board has a toolbar and a canvas. Users can create and edit a knowledge graph on the canvas. This is an example: \textit{``each cluster belongs to a zone''}. (A) an entity card (attributes and query), (B) a relation card (semantic and query), and (C) the query template entry.}
  \label{fig:building_board}
\end{figure}
\subsubsection{Building board}
The building board supports the creation of knowledge graphs.
Users can add entities and relations freely to the building board (see~\cref{fig:building_board}).
Both entities and relations are displayed as cards.

\textbf{\textit{Entities and attributes (\cref{fig:building_board}A).}}
Each card has a title showing the name of the entity, a two-column table about the attributes of the entity, and a button to add new attributes.
Each row of the two-column table shows the index, the name, and the data type of an attribute.
Possible data types are introduced in~\cref{sec:data}.
Attributes can be edited by directly double-clicking the corresponding texts.
Besides, users can predefine the query template after clicking the edit button (\cref{fig:building_board}C).
The format of query templates follows the grammar of Kusto.

\textbf{\textit{Relations (\cref{fig:building_board}B).}}
Each relation is represented by an arrow link connecting two entities, and a label with the name of the relation in the middle of the arrow.
Users may toggle the label of the relation to fold or unfold the detailed information of the relation, which is shown as relation cards. 
Similar to the entity card, users are allowed to edit the semantic and query template of the relation link.

\textbf{\textit{Interactions.}}
Users are allowed to import existing knowledge graphs or store the knowledge graph in JSON form for later use via clicking corresponding buttons in the menu bar.

\subsection{Monitor Anomalies}
The overview serves as an outline of anomalies in cloud computing systems (\textbf{R2}). 
It mainly consists of two panels: the Incident Log Panel on the left side and the KPI Panel on the other. 
Since it is time-consuming to query incident logs and KPIs online from a cloud-distributed database, RCInvestigator stores caches for users to speed up the process. 
Users can create a new cache by clicking the \textit{``New Cache''} button and configure the query template in an opened drawer. 
Also, users can select a cache from the cache list and click the \textit{``Start''} button to view the incident logs and KPIs.

\textbf{\textit{The incident log panel.}}
This panel lists incident logs during a selected period of time in table form to provide necessary clues for further investigation. 
Each row represents an incident log and each column means an attribute of incidents. 

\textbf{\textit{The KPI panel.}}
This panel demonstrates the main KPIs of pre-selected zones.
These KPIs and zones are configured when a user creates a data cache.
The data types of KPIs are the same as attributes.
RCInvestigator visualizes KPIs by aligned line plots and Gantt plots.

\textbf{\textit{Interactions.}}
There are two types of interactions.
One is when a user double-clicks an incident log, the corresponding KPI plot will be highlighted.
The other is when a user can brush-select a time range with anomalies and then enter the next reasoning stage.

\subsection{Investigate Root Causes}
To facilitate root cause reasoning, RCInvestigator integrates novel hypothesis interactions and intuitive visualizations (\textbf{R3}).
For each hypothesis interaction, a module of the reasoning model is designed to support collecting and recommending relevant clues (\textbf{R4}).
All reasoning steps are completed on the reasoning board.
We introduce the visual designs and interactions as well as the corresponding reasoning model.

\begin{figure}[t]
  \includegraphics[width=\linewidth]{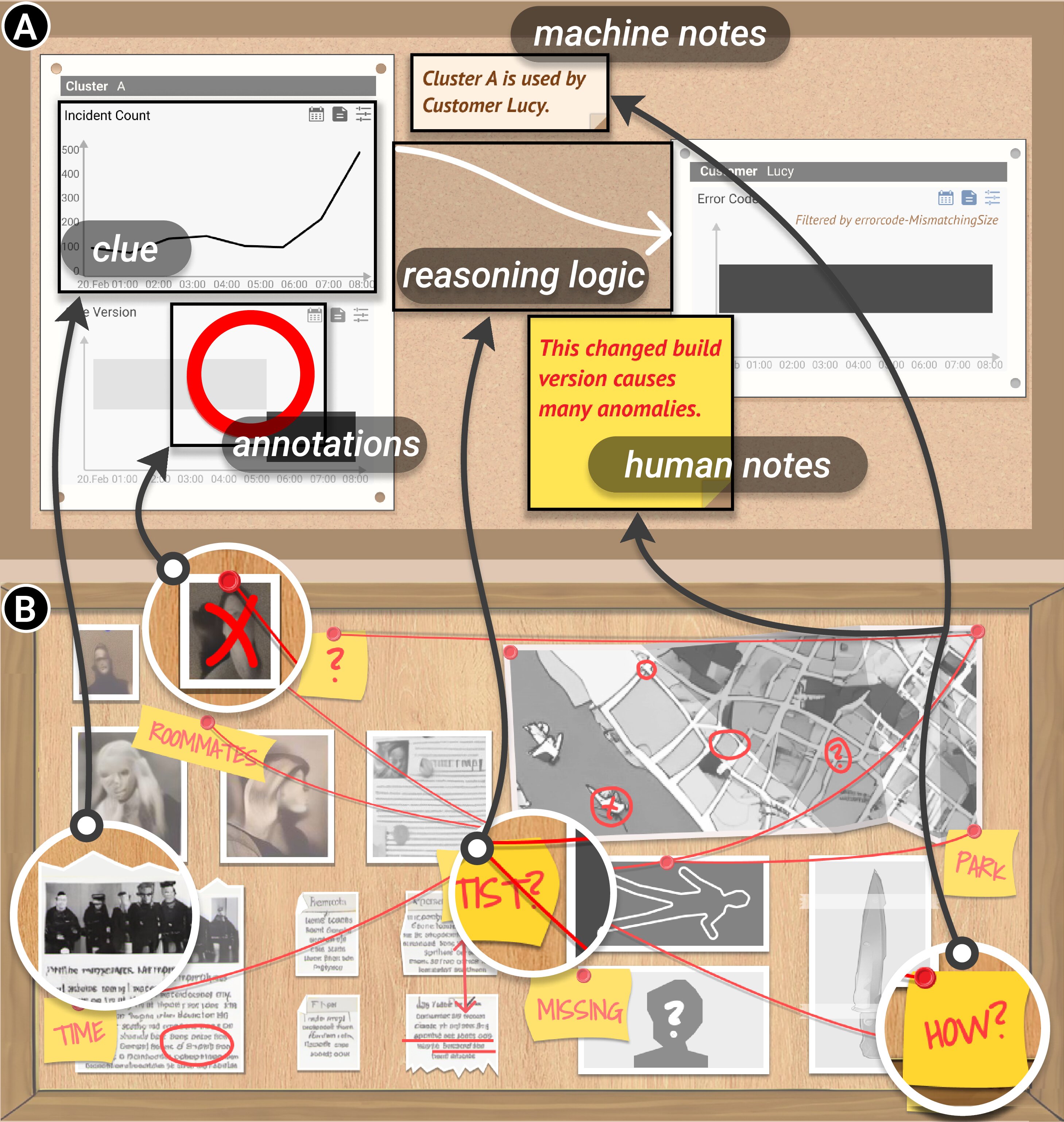}
  \caption{This shows the metaphor mapping of our design. (A) is the investigation board in RCASleuth, while (B) is the investigation board in the real world. There are four types of typical elements, including clues, reasoning logic, annotations, and notes. In RCASleuth, we add the notes from the machine agent.}
  \label{fig:investigation_board}
\end{figure}

\textbf{\textit{Visual designs.}}
The investigation board not only features investigation elements, including cause clues and reasoning logic but also serves as an interactive medium for human-machine collaboration. 
Our design for the investigation board is inspired by the investigation board used by detectives to explore the causes of accidents in the real world. 
Detectives collect various clues and display them on cards on the board, then connect the clues with red lines and notes (\cref{fig:investigation_board}B). 
Similarly, our design is clue-linked to form a cohesive understanding of the investigation (\cref{fig:investigation_board}A).
Each clue is presented as a simple plot.
Since the data type of clue is diverse, we adopt different visualizations.
For data types including string and set, the Gantt graph is employed to visualize event sequences.
For data types including number and bag, the line plot is used to display time series.
All attribute cards are organized into different larger entity cards that group clues related to the same entity.
Every entity card has the title ``Entity Concept: Entity'', like \textit{``Area: Asia''}.
Besides the cause clues, reasoning logic is visualized as the arrow link with machine notes, connecting entity cards.

\textbf{\textit{Interactions overview.}}
RCInvestigator supports three categories of hypothesis interactions: \textit{expanding} and \textit{refining} for clues as well as \textit{annotating} for reasoning logic.
 The \textit{expanding} is designed for extending the collection size of alternative cause clues and gaining generalized clues; 
 The \textit{refining} is designed for making a specific clue detailed.
 The \textit{annotating} of reasoning logic can be collaboratively completed by humans and machines to illustrate the connection between clues and high-level interpretation of visual patterns.
Next, we introduce interactions and how the model supports them in detail.

\subsubsection{Expand Relevant Clues}
This section presents five expanding hypothesis interactions, an expanding model, and a layout algorithm.
The expanding model supports hypothesis interactions.
Besides, the layout algorithm keeps the semantic layout of clues as well as reduces visual clutters.

\textbf{\textit{Interactions and visual designs.}} 
Five expanding hypothesis interactions are intended to make it easier for users to organize their thoughts.
Each interaction serves a specific expansion purpose and corresponds to a certain expanding direction.
Specifically, these interactions include: upward, downward, leftward, rightward, and inward.
They are summarized from the knowledge graph model.
We model a clue as a tuple $c_{i_p}$ = ($e_i$, $a_{i_p}$).
To make the expansion systematic and reasonable, each expansion interaction only changes one dimension of the tuple.
The in-ward interaction expands the attribute $a_{i_p}$ while up-, down-, left-, and right-ward interactions expand the entity $e_i$.

\begin{wrapfigure}[2]{l}{0.2cm}
\centering
\includegraphics[width=0.5cm]{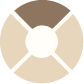}
\vskip-\intextsep
\end{wrapfigure}
\noindent \textit{Upward expansion ($f_U$)} allows users to collect clues from entities of a higher hierarchy.
This interaction is useful when users need to generalize some root causes. 
For example, if a user is collecting clues about a specific cluster, the upward expansion could provide clues about the cluster's zone. 
This can give users a more comprehensive understanding of the root cause, especially the possibly affected range. 

\begin{wrapfigure}[2]{l}{0.2cm}
\centering
\vskip-\intextsep
\includegraphics[width=0.5cm]{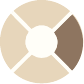}
\vskip-\intextsep
\end{wrapfigure}
\noindent \textit{Rightward expansion ($f_R$)} allows users to collect clues from entities in the same hierarchy but of different categories.
This interaction is useful when users need to collect diverse clues.
For example, if a user finds that a customer encounters many incidents, the rightward expansion could provide clues about the allocations required by the customer. 
This helps to expand the possible cause scope via relations.

\begin{wrapfigure}[2]{l}{0.2cm}
\centering
\vskip-\intextsep
\includegraphics[width=0.5cm]{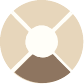}
\vskip-\intextsep
\end{wrapfigure}
\noindent \textit{Downward expansion ($f_D$)} allows users to collect clues from entities of a lower hierarchy.
This interaction is useful when users validate clues. 
For example, if a user finds many incidents of the same error occurring in a zone, the downward expansion could provide clues of a detailed incident on a certain cluster. 
This would help the user gain deeper insights because the same error might be caused by different root causes and a detailed comparison could help identify the differences.

\begin{wrapfigure}[2]{l}{0.2cm}
\centering
\vskip-\intextsep
\includegraphics[width=0.5cm]{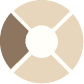}
\vskip-\intextsep
\end{wrapfigure}
\noindent \textit{Leftward expansion ($f_L$)} allows users to collect clues from sibling entities, which have the same entity concept but are different instances.
This interaction is particularly useful when users need to compare siblings or find analogous causes.
For example, if a user finds a root cause that affects clusters in the US, the leftward expansion could provide similar clues in Canada. 
This would help the user locate analogous causes more quickly.

\begin{wrapfigure}[2]{l}{0.2cm}
\centering
\vskip-\intextsep
\includegraphics[width=0.5cm]{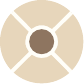}
\vskip-\intextsep
\end{wrapfigure}
\noindent \textit{Inward expansion ($f_I$)} allows users to expand clues by adding new attributes of the same entity.
This interaction is particularly useful when users need to find clues that are related to a different aspect of the same entity. 
For example, if a user collected clues like the stability of a specific cluster, the inward expansion could provide other attributes like utilization. 
This would help the user collect clues from more aspects and find complex root causes.

\textbf{\textit{The expanding model.}}
To support expanding interactions, we design a model to collect clues from various directions on the constructed knowledge graph. 
The data types of clues are diverse, so we first define the relevance between two clues before introducing the search steps.
Based on the observation that the closer and more overlapping the change points of two clues, the greater the relevance between them, we calculate the change points via the popular BCPD~\cite{bocp}.
Then, the distance between two arrays of change points $S$ and $T$ is defined in Eq.~\ref{eq:dis}.
For each point $x \in S$, we find the nearest point $y(x)\in T$ and calculate their distance $|x-y(x)|$. 
\begin{equation}
    d(S, T) = \sum_{x\in S, y(x) \in T}|x-y(x)|
\label{eq:dis}
\end{equation}
Assuming that the change point sequences of clues $c_{1}$ and $c_{2}$ are $S_{1}$ and $S_{2}$, respectively.
The relevance is calculated via \cref{eq:rel}
\begin{equation}
    R(c_{1}, c_{2}) = 1 - \frac{1}{2N}(\frac{d(S_{1},S_{2})}{|S_{1}|} + \frac{d(S_{2},S_{1})}{|S_{2}|})\\
\label{eq:rel}
\end{equation}
, where $1/2N$, $|S_{1}|$, and $|S_{2}|$ are normalization terms.

The search process for inward expansion and the other four are different.
Inward expansion has two steps:
(1) For the current clue $c_{ip}$ = ($e_i$, $a_{ip}$), the model queries for other attributes of entity $e_i$ and records queried attributes except added ones.
(2) It computes the relevance between attributes and the current one, reserving new clues with top 5 relevance.
The other four expansions have three steps:
(1) For the current clue $c_{ip}$, the model searches for one-hop entities via relations.
For example (see Fig.~\ref{fig:build}), the one-hop entities of \textit{``Zone''} include \textit{``Area''}, \textit{``Cluster''}, and \textit{``Customer''}.
(2) The model queries for attributes of those one-hop entities and computes the relevance between alternative clues and the current clue.
The model keeps a result list with the top 5 relevance in each direction so it updates the list if the new alternative clue has a higher relevance.
(3) If there is at least one new clue of an entity updating the result list, the model executes one-hop expansion again for such an entity.
For example (see Fig.~\ref{fig:build}), if a clue from entity \textit{``Customer''} is added to the result list, the new expansion will reach \textit{``Allocation''}.
Specifically, visited entities will not be expanded again.
The model repeats steps one and two until it reaches the time limit (2s) or there is no higher relevance new clue.

\textbf{\textit{Layout.}}
After many clues are added to the panel, organizing and presenting these clues in an organized way while retaining directional semantics is important.
We propose a two-step layout algorithm based on DAGre~\cite{dagre}.
First, we divide entity cards into different groups and calculate the layout in each group.
Two entity cards belong to the same group if there is an up- or down-ward reasoning logic link between them.
For each group, we use DAGre to layout entity cards vertically and define the ($x_1$,$y_1$) of each entity card.
Second, we calculate the layout of all groups.
In this step, each group is treated as one card and organized horizontally.
We calculate the coordinates ($x_2$,$y_2$) of each group.
The final coordinates of each entity card are ($x_1 + x_2$, $y_1 + y_2$).

\textbf{\textit{Justification.}}
To design a reasonable expansion interaction, we conducted a literature review.
We observed the database query scripts used by experts in their daily research processes. 
We found that most existing collection methods are designed from a model perspective, but they neglect the user agent's understanding difficulty. 
Existing models include enumeration-based methods that compute a large number of KPI associations and collect clues with the highest correlation, as well as search-based methods that find adjacent nodes on a constructed graph. 
Although these methods can find relevant clues, they do not systematically organize clues based on semantic meaning, so users cannot freely control the direction of the collection. 
Therefore, to overcome this difficulty, we hope to summarize a semantic-level expansion interaction for the user agent, which can help users communicate with the model from a higher semantic level to collect clues in a more organized way.

\subsubsection{Refine Current Evidence}
After collecting clues, it is necessary to refine them to improve the accuracy of the reasoning process. 
This is important because anomalies in the same period can be caused by different factors and need to be investigated separately. 
For example, in a cascading anomaly scenario, a patch designed for operating system A (OS-A) may trigger inconsistency failures (Error 1), causing users to frequently apply for resources with OS-A and leading to network congestion (Error 2).
In this case, analysts may only observe one peak in the KPI but need to investigate different error types and operating systems (Error 1 with OS-A, Error 2 with other OSs).
Additionally, different groups of filters can reveal diverse insights and patterns.
For instance, incidents may only occur on resources of a certain size and version.
To address this, RCInvestigator supports refining clues through detailed filters, such as error types, and mines valuable filters to identify the root cause. 
We propose the filtering interaction, design a filter card, and adapt a refining model.

\begin{figure}[ht]
  \includegraphics[width=\linewidth]{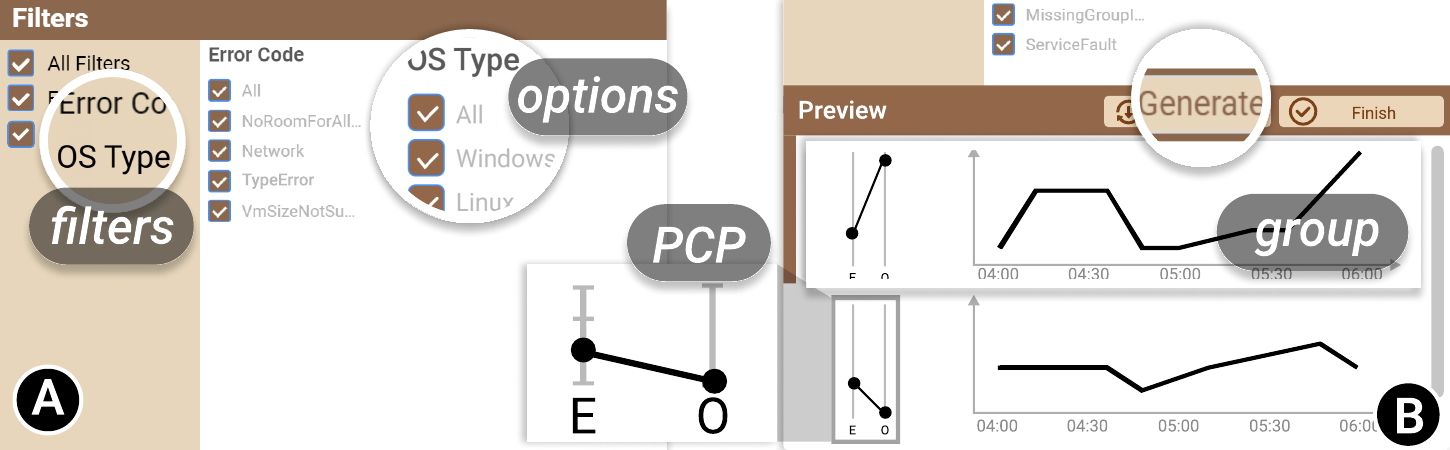}
  \caption{The filter card consists of two parts: (A) users can select filters and options in the selection panel; (B) the preview panel displays alternative groups of options and filtered attributes. For example, the PCP shows \textbf{E}rrorcode-TypeError (3$^{rd}$) with \textbf{O}SType-Linux (2$^{nd}$).}
  \label{fig:filter}
\end{figure}

\textbf{\textit{The filter card} (Fig.~\ref{fig:filter})} allows users to select desired filter types and displays previews of attributes after implementing selected filter groups.
The filter card consists of two parts: the selection panel (Fig.~\ref{fig:filter}A) and the preview panel (Fig.~\ref{fig:filter}B).
In the selection panel, all filters are listed on the left side.
When a user selects a filter like OS type, corresponding options like Linux will be displayed on the right side.
The user can check or uncheck the options as needed. 
The combinations of options are diverse, but only those that reveal a similar trend are effective. 
For example, while Error1 with OS-A reveals an increase in incident numbers, Error2 with OS-A may not be significant. 
The user can click ``generate'' to let the refining model search for effective combinations. 
The preview panel displays option combinations and corresponding filtered attributes.
We visualize the option combination by parallel coordinate plots and attributes by Gantt plots or line plots.

\textit{\textbf{The refining model}} searches for combinations of filter options that make the current attribute most relevant to existing clues.
We review many models that can automatedly extract main trends of attributes by mining valuable groups of filters.
One of the latest and most relevant tools is MID~\cite{aidice}.
It adopts a meta-heuristic search to find option combinations that maximize the increasing trend of the incident number attribute.
We modify it to adapt our problem in two folds.
First, we change the objective function.
We implement the search twice for both minimize and maximize because we want to find corresponding filter combinations for both main decreasing and increasing trends, respectively. 
Second, we change the filter selection scope, letting it only search for combinations involving pre-selected filters.
After the combination is found, we compute the average relevance between the filtered attribute and add clues via Eq.~\ref{eq:rel}.

\subsubsection{Annotate Reasoning Logic}
Reasoning logic is necessary to assist analysts in interpreting the investigation process.
In RCInvestigator, the reasoning logic proposed by humans or machines is annotated differently.

\textbf{\textit{Machine annotations.}}
In the reasoning process, the machine agent annotates two categories of logic: (1) summaries of collected clues and (2) annotations of filtered attributes.
These two categories are generated automatically by predefined templates.
The collection logic is
\begin{center}
\textit{<$e_i$><$e_{i_1}$>...<$e_{i_p}$>[semantic]<$e_j$><$e_{j_1}$>...<$e_{j_q}$>}
\end{center}
Specifically, the \textit{[semantic]} is predefined in the query template of relation.
For filtered attributes, the template is
\begin{center}
\textit{Filtered by [<$b_i$><$c_{i_1}$>..<$c_{i_k}$>]$^+$}    
\end{center}
All selected filters and options will be spanned in the annotation.
To ensure correct grammar, especially the form of verbs, we adopt the package compromise~\cite{compromise} to eliminate the typos.

\textbf{\textit{Human annotations.}}
RCInvestigator allows users to add shapes and texts freely to the reasoning board.
This is to make the reasoning logic expressive and flexible.
With shapes, including circles, rectangles, and arrows, users can highlight anomalies and connect them on demand.
Moreover, texts with different colors support distinguishing hypotheses, conclusions, and mitigation suggestions.

\subsection{Output Results}
After users find the root cause and make annotations, they may click the camera icon to save the analysis process as a PNG image, which will help users share the results through discussion boards or email. Users can also download the anomaly analysis log in JSON form and share it with others for better collaborative analysis.
\section{Case Study}

\subsection{Use Case 1: Updated Inconsistent Strategy}
EA received an alert email: there were plenty of incidents happening in \textit{Area01}.
EA investigated the root cause of these incidents.

\begin{figure}[ht]
  \includegraphics[width=\linewidth]{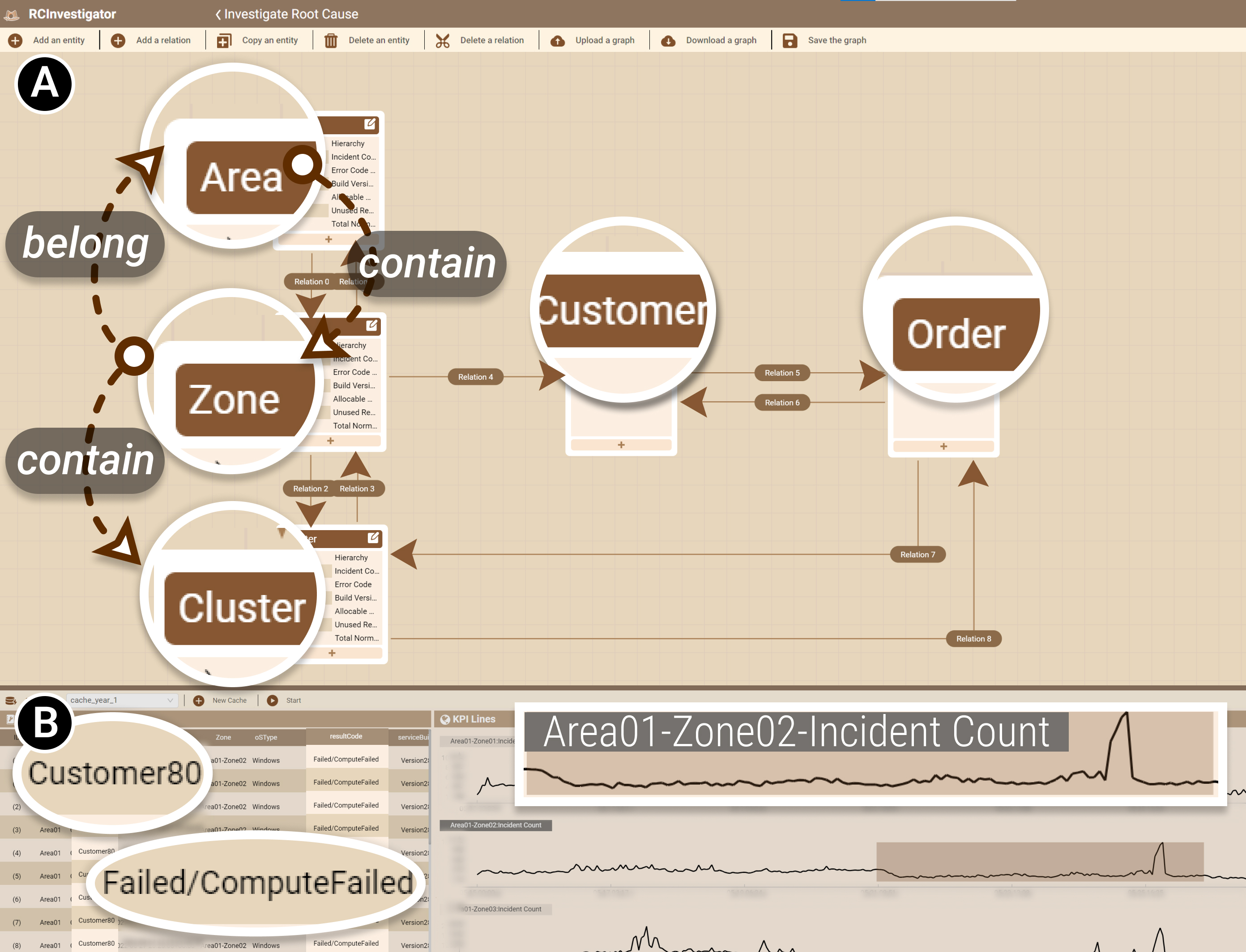}
  \caption{The first and second steps of Case 1.
  (A) EA built a knowledge graph with 5 entities and 9 relations based on her domain knowledge, such as each region contains many Zones.
  (B) EA observed many incidents were related to \textit{Customer80} and happened in \textit{Area01-Zone02}.
  }
  \label{fig:build}
\end{figure}

\textbf{S1. Build a knowledge graph (Fig.~\ref{fig:build}A).}
EA first created an entity and named it \textit{Area}, which is the highest level of the physical structures.
Then she added necessary attributes like \textit{Incident Count (the number of hourly incidents)} and \textit{Unuse Reserved VMs (the number of virtual machines reserved by customers but unused)}.
EA wrote a query template to define how to retrieve such attributes from databases.
Similarly, EA duplicated an entity \textit{zone}.
Next, EA connected \textit{zone} and \textit{area} with a relation, as an area contains many zones.  
EA wrote a query template to constrain how to find related zones of an area.
Finally, EA repeatedly created entities and relations and finished the whole knowledge graph, which is composed of five entities and nine relations.
EA downloaded and saved the knowledge graph.

\textbf{S2. Find that many incidents happen in a short period (Fig.~\ref{fig:build}B).}
EA selected a cache of the database and quickly scanned recent incident logs. 
She found that a single user, \textit{Customer80}, had encountered many incidents logged as \textit{Failed/ComputeFailed} in a short period. 
EA double-clicked on the first incident log and saw that the corresponding KPI line \textit{Area01-Zone02-Incident Count} was highlighted. 
She then brush-selected a time range of nearly one week, containing peak values.

\begin{figure*}[ht]
  \includegraphics[width=\linewidth]{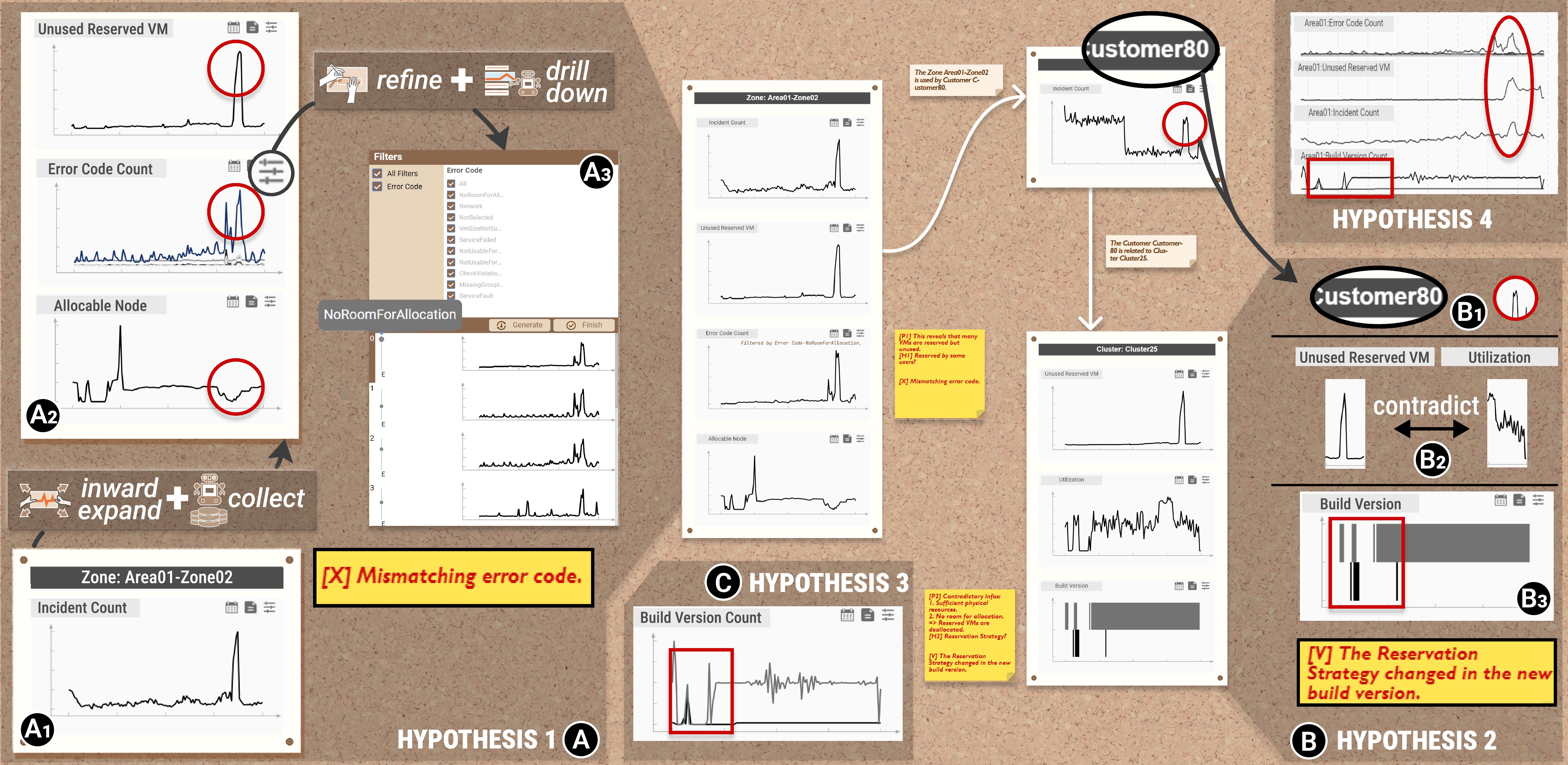}
  \caption{The investigation process of Case 1.
  EA proposed four hypotheses.
  (A) In Hypothesis 1, EA used inward expansion to collect three clues and refined the error code clue. She negated this due to the mismatching error code.
  (B) In Hypothesis 2, EA collected clues and found a contradicting pattern.
  She confirmed this due to a change in the build version.
  Both (C) and (D) display the influence range of such a root cause.
  }
  \label{fig:case1}
\end{figure*}

\textbf{S3. Investigate why there are a large amount of incidents.}
An entity card appeared on the investigation board, surrounded by recommended clues in five directions. 
EA was drawn to the inner attributes of the current entity, \textit{Area01-Zone02}. 
She noticed that the number of \textit{unused reserved VMs} and \textit{error code count} peaked while \textit{allocable nodes} decreased (Fig.~\ref{fig:case1}A$_2$) at nearly the same time as \textit{the incident count} (Fig.~\ref{fig:case1}A$_1$). 
This phenomenon led to many possible root causes. 
Her first hypothesis (Fig.~\ref{fig:case1}A) was that several users reserved many allocable nodes in a short period, causing a jam. 
To further examine this hypothesis, EA checked the \textit{error code} (Fig.~\ref{fig:case1}A$_3$). 
She clicked on the filter to preview the main error code count.
If her guess was correct, the main code should have been related to configurations. 
However, she found that the main error code was \textit{No Room for Allocation}, which did not match her guess.
Due to the mismatching error code, EA negated her first hypothesis about the customers.

EA then made the second hypothesis (Fig.~\ref{fig:case1}B) that the root cause might be related to the reservation process as it appears that many VMs have been reserved without as many reservation requests from customers.
Specifically, EA thought there might be some VMs that were not reserved and released correctly.
She decided to investigate a customer who met this incident for further insight, so EA added \textit{Customer80} (Fig.~\ref{fig:case1}B$_1$).
Then, because some VMs in \textit{Cluster25} were reserved by \textit{Customer80} she also added it. 
In \textit{Cluster25}, she found a peak in the \textit{unused reserved VMs}.
Besides, EA observed that the \textit{utilization} (Fig.~\ref{fig:case1}B$_2$) was low, indicating there were sufficient physical resources.
In contrast, the \textit{error code} was \textit{No Room For Allocation}, which meant there were no available nodes on physical resources for allocation. 
These contradictory facts made her further confirm the second hypothesis. 
Then, a change in the \textit{build version} (Fig.~\ref{fig:case1}B$_3$) about two days before the incidents caught her attention. 
EA found that this latest update included a reservation strategy, leading her to conclude that the root cause was hidden behind the new build version.

Next, EA explored the \textit{build version count} in \textit{Area01-Zone02} to investigate the influence range of the root cause (Fig.~\ref{fig:case1}C). 
She found that the latest build affected the entire \textit{Zone02}, as the \textit{build version count} also changed a few days before. 
Additionally, a similar pattern occurred in the whole \textit{Area01} (Fig.~\ref{fig:case1}D). 
In summary, EA determined that the root cause was the reservation strategy.

\textbf{S4. Annotate and save.}
Having gathered enough clues to verify her reasoning, EA began making annotations to record her thoughts. 
She circled anomalous evidence such as the peak in \textit{allocable nodes} and \textit{unused reserved VMs}, used rectangles to highlight key clues that supported or negated hypotheses, and completed her annotations. 
EA negated her first hypothesis about \textit{customers} due to the mismatching \textit{error code}. 
Her second hypothesis was supported by contradictory factors and changed \textit{build versions}, while her third inference about influence range was confirmed by similar patterns in \textit{Area01}. 
Finally, she summarized the true root causes and provided mitigation suggestions. 
EA exported the investigation process as an image for sharing to report the incidents and the process of root cause analysis.

\subsection{Use Case 2: Normal Nodes Error}
The alert email EC received informed him that the allocable nodes in \textit{Area07-Zone01} decreased dramatically.
\begin{figure*}[ht]
  \includegraphics[width=\linewidth]{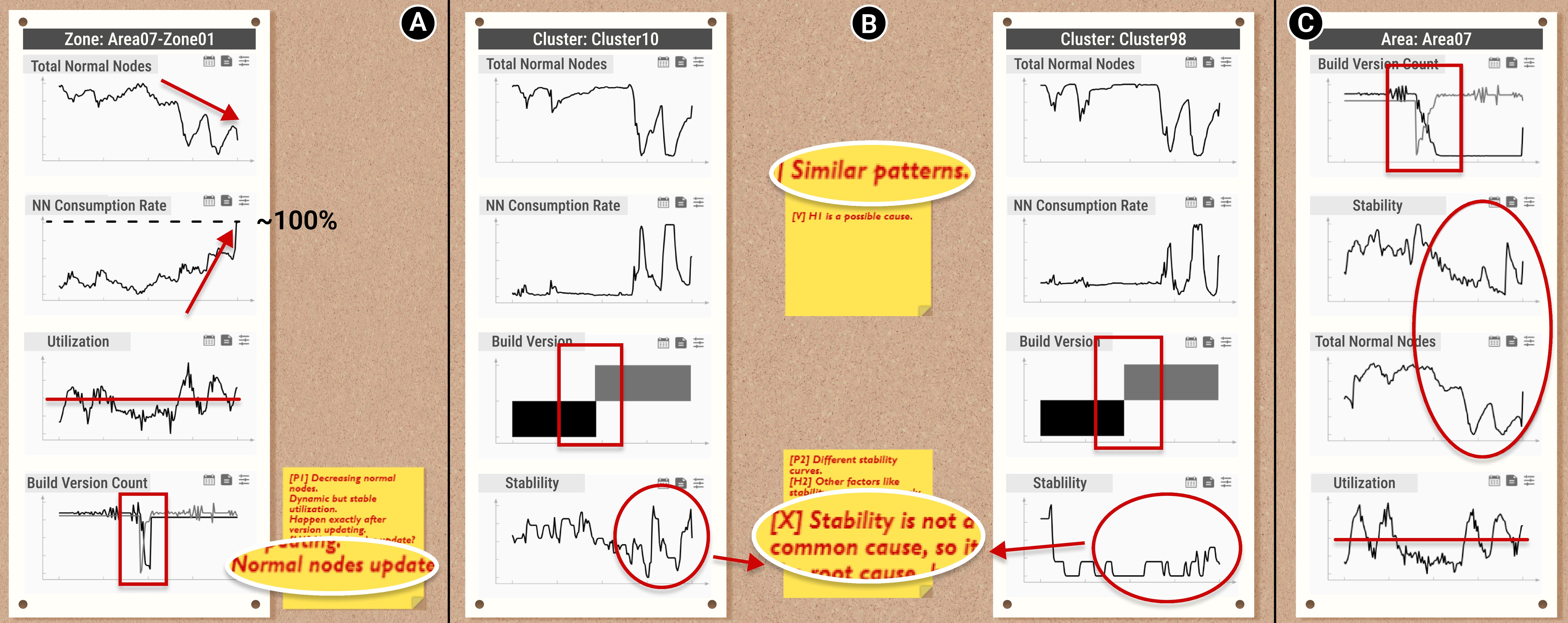}
  \caption{The investigation process of Case 2.
  EC investigated clues from three different hierarchies.
  (A) EC found a drop in normal nodes and a high consumption rate.
  He thought it might be caused by a recent update.
  (B) EC further checked his guess: excluded the stability factor on clusters and confirmed the build version update.
  (C) EC investigated the influence range of the root cause.
  }
  \label{fig:case2}
\end{figure*}

\textbf{S1\&2. Find that the number of normal nodes drops dramatically.} 
EC loaded the knowledge graph built by EA earlier and selected a data cache. 
After observing the KPI lines for a few seconds, EC found that the number of \textit{total normal nodes} in \textit{Area07-Zone01} had been dropping for a few days, and the trend had become steeper over time. 
Therefore, EC brush-selected the time range for further analysis.

\textbf{S3. Investigate why the normal nodes decrease.}
EC found that while the total number of \textit{normal nodes} was decreasing, the \textit{consumption rate} of normal nodes was consistently increasing and approaching 100\% (Fig.~\ref{fig:case2}A). 
However, the \textit{utilization} has not shown an overall upward or downward trend in its periodic fluctuations. 
This was highly illogical and suggested that once the available normal nodes were consumed, they were not being utilized as they should be. 
Moreover, EC discovered that the \textit{build version} update was almost simultaneous with the start point of the upward trend in \textit{total normal nodes}. 
Thus, EC suspected that this update may have caused the current issue, possibly due to a problem with the recycling or validating of normal nodes.

However, EC still needed to check cluster instances and exclude any stability or other factors that might be causing the issue (Fig.~\ref{fig:case2}B). 
Thus, EC added \textit{Cluster10} and \textit{Cluster98}, two of the most relevant clusters, and found that their \textit{consumption rate}, \textit{utilization}, and \textit{build version} were very similar.
However, the \textit{stability} changed in \textit{Cluster10} while remaining balanced in \textit{Cluster98}. 
Since the change in \textit{stability} was not a common pattern, EC tentatively concluded that the build version update affecting normal nodes might be the true cause of the issue.
Furthermore, EC also discovered that the update had covered the entire \textit{Area07} (Fig.~\ref{fig:case2}C) as there were similar patterns.

\textbf{S4. Conclude the summary of this investigation.}
Finally, EC added his reasoning logic. 
He wrote down three hypotheses and highlighted peaks and drops. 
Additionally, EC summarized the possible solution, which was to investigate the code blocks related to \textit{normal node} judgment and recycling in the build version update. 
EC then exported this analysis report and completed the root cause analysis.

\subsection{Expert Interview}
We collected feedback from four domain experts (EA, EB, EC, and ED) via one-on-one structured interviews.
The whole process consists of three stages: (1) \textbf{[10min]} introduction of RCInvestigator; (2) \textbf{[50min]} free investigation of anomalies' root causes; (3) \textbf{[10min]} a structured interview.
We summarized the feedback from three aspects as follows.

\textbf{Investigation Framework.}
All experts praised the proposed investigation framework.
EA said the building stage facilitated investigation experience forming and sharing and was useful for formulating knowledge basis.
Besides, EB tried to edit existing investigation knowledge and said it was easy to extend.
EC and ED both commented that RCInvestigator enabled easy reuse of investigation experience.
Moreover, all experts highlighted that RCInvestigator reduced the laboring effort needed in data collection.
It is effective as they do not need to switch between databases and analysis panels during the investigation, so the investigation becomes more \textit{``fluent''} and \textit{``coherent''}.

\textbf{Visualization and Interactions.}
Experts agreed that visualizations and interactions adopted in RCInvestigator were intuitive and helpful.
All experts said that the semantic layout of clues helped them organize their minds logically as well as inspire further investigation.
Besides, EA and ED praised the collaborative summaries and annotations.
EA said, \textit{``Direct conclusions given by the machine agent help me value clues quickly.''}
Moreover, all experts commented that the design was expressive and easy-to-understand.
\textit{``Through simple charts, I can fully understand what happened and thus devote my effort all in the investigation instead of reading the chart''}.

\textbf{Suggestions.}
Experts also give many valuable suggestions.
EA said that we should allow deleting unexpected clues.
EB and EC suggested adding a mini-map of the knowledge graph during the investigation because it provides an overview and facilitates a comprehensive understanding of the whole reasoning.
ED said we could add a dashed line while hovering on charts as it would be easier to align time across charts.
We improved RCInvestigator based on these suggestions.
\section{Discussion}
This section presents the significance and generalizability, lessons learned, limitations, and future work of RCInvestigator.

\textbf{Significance and generalizability.}
We discuss the significance of RCInvestigator in the following two aspects.
    
    \textit{Investigation framework.} We proposed a root cause investigation framework based on human-machine-collaborative schema. 
    With this framework, analysts can be freed from labor-intensive tasks such as data querying and information gathering, and instead focus on thinking, reasoning, and analysis. 
    Meanwhile, machines collect data, recommend clues, and summarize information controllably. 
    Though this framework is designed for cloud computing systems, we argue that it can be applied to other scenarios, like smart manufacturing RCA.
    Especially in scenarios where RCA requires analyzing large amounts of data from multiple sources and heavily relies on domain knowledge.

    \textit{Techniques.} RCInvestigator comprises a novel interactive root cause reasoning model and a real-world investigation board-inspired time-oriented data visualization, which helps to identify the complex underlying causes behind anomalies.
    This paper proposes two types of hypotheses interactions and corresponding models for time-oriented data exploration.
    These interactions and models can also be extended to other scenarios involving time-oriented data.

\textbf{Lessons Learned.}
Through our collaboration with experts in developing RCInvestigator, we have learned valuable lessons. 
First, providing necessary textual summaries when recommending visual elements is crucial. 
In our initial design, all recommended clues are presented via small multiple visualizations in a single panel with the descending order of association.
However, experts provided feedback that even the simplest visual patterns can be overwhelming when they appear in large numbers. 
Instead, if recommended visual patterns can be classified and summarised in natural language, it can effectively assist in understanding the recommendations. 
Second, layouts with semantic meaning can more efficiently present information. 
Initially, we used force-directed graph layout algorithms to reduce crossings when designing expanded interactions using the steering wheel metaphor. 
However, we later discovered that even though force-directed graphs can effectively reduce visual clutters and crossings, this layout is difficult for users to understand.
Specifically, users usually get lost in the investigation and spend much time deciding to expand which clue.
However, by using a layout with semantic meaning that aligns with users' thought patterns, which preserves the semantics of the five directions, it is easier to maintain immersive analysis and exploration.

\textbf{Limitaions and future work.}
RCInvestigator has some limitations that should be considered.  
First, the current investigation knowledge building is human-dominated, which ensures reliability but can be labor-intensive.
While automated knowledge extraction techniques exist, they are limited by a lack of labeled data so they may fail to extract high-level investigation knowledge. 
This can increase the query cost of the entire reasoning process.
Therefore, we leave the intelligent building approach as future work.
Second, the evaluation of RCInvestigator is currently limited due to challenges related to data compliance and analyst recruitment.
We only conduct preliminary validation of RCInvestigator through case studies and expert interviews.
In future work, we plan to conduct additional field studies and fully integrate RCInvestigator into cloud computing systems.

In addition, we identify expansion potential for RCInvestigator. 
First, RCInvestigator is based on mouse and keyboard interaction to allow users to propose hypotheses, but we plan to allow users to propose hypotheses in more advanced ways in the future, such as using natural language. 
Second, template-based annotation methods can be improved.
We plan to automatically summarize insights and support multimodal annotation. 
Third, we plan to expand RCInvestigator to support multiple analysts to collaborate simultaneously.
\section{Conclusion}
In this paper, we proposed RCInvestigator, an interactive system for investigating anomaly root causes in cloud computing systems. 
Through the collaboration with domain experts, three challenges were identified in investigating root causes.
We designed a novel human-machine-collaborative framework consisting of four stages (building, monitoring, reasoning, and concluding) to address these challenges.
We built RCInvestigator based on this framework and evaluated the system through two real-world use cases, receiving positive feedback from experts.
\bibliographystyle{abbrv-doi-hyperref}

\bibliography{template}

\begin{thebibliography}{10}

\bibitem{alog}
P.~Aggarwal, A.~Gupta, P.~Mohapatra, S.~Nagar, A.~Mandal, Q.~Wang, and A.~Paradkar.
\newblock {Localization of operational faults in cloud applications by mining causal dependencies in logs using golden signals}.
\newblock In {\em {Proceedings of the International Conference on Service-Oriented Computing (Workshop)}}, pp. 137--149, 2020. \href{https://doi.org/10.1007/978-3-030-76352-7_17}
{doi: {{%
10\hspace{.1pt}\discretionary{.}{%
}{.}\hspace{.4pt}1007\discretionary{/}{%
}{/}978\discretionary{%
}{-}{-}3\discretionary{%
}{-}{-}030\discretionary{%
}{-}{-}76352\discretionary{%
}{-}{-}7\_17}}}


\bibitem{AIGNER2007401}
W.~Aigner, S.~Miksch, W.~M\"{u}ller, H.~Schumann, and C.~Tominski.
\newblock Visualizing time-oriented data — {A} systematic view.
\newblock {\em Computers \& Graphics}, 31(3):401--409, 2007. \href{https://doi.org/10.1016/j.cag.2007.01.030}
{doi: {{%
10\hspace{.1pt}\discretionary{.}{%
}{.}\hspace{.4pt}1016\discretionary{/}{%
}{/}j\hspace{.1pt}\discretionary{.}{%
}{.}\hspace{.4pt}cag\hspace{.1pt}\discretionary{.}{%
}{.}\hspace{.4pt}2007\hspace{.1pt}\discretionary{.}{%
}{.}\hspace{.4pt}01\hspace{.1pt}\discretionary{.}{%
}{.}\hspace{.4pt}030}}}


\bibitem{aigner2023visualization}
W.~Aigner, S.~Miksch, H.~Schumann, and C.~Tominski.
\newblock {\em {Visualization of Time-Oriented Data}}.
\newblock 2023. \href{https://doi.org/10.1007/978-1-4471-7527-8}
{doi: {{%
10\hspace{.1pt}\discretionary{.}{%
}{.}\hspace{.4pt}1007\discretionary{/}{%
}{/}978\discretionary{%
}{-}{-}1\discretionary{%
}{-}{-}4471\discretionary{%
}{-}{-}7527\discretionary{%
}{-}{-}8}}}


\bibitem{timecurve}
B.~Bach, C.~Shi, N.~Heulot, T.~Madhyastha, T.~Grabowski, and P.~Dragicevic.
\newblock Time {C}urves: Folding time to visualize patterns of temporal evolution in data.
\newblock {\em IEEE Transactions on Visualization and Computer Graphics}, 22(1):559--568, 2016. \href{https://doi.org/10.1109/TVCG.2015.2467851}
{doi: {{%
10\hspace{.1pt}\discretionary{.}{%
}{.}\hspace{.4pt}1109\discretionary{/}{%
}{/}TVCG\hspace{.1pt}\discretionary{.}{%
}{.}\hspace{.4pt}2015\hspace{.1pt}\discretionary{.}{%
}{.}\hspace{.4pt}2467851}}}


\bibitem{patient}
T.~Baumgartl, M.~Petzold, M.~Wunderlich, M.~Hohn, D.~Archambault, M.~Lieser, A.~Dalpke, S.~Scheithauer, M.~Marschollek, V.~M. Eichel, N.~T. Mutters, H.~Consortium, and T.~V. Landesberger.
\newblock In search of patient zero: Visual analytics of pathogen transmission pathways in hospitals.
\newblock {\em IEEE Transactions on Visualization and Computer Graphics}, 27(2):711--721, 2021. \href{https://doi.org/10.1109/TVCG.2020.3030437}
{doi: {{%
10\hspace{.1pt}\discretionary{.}{%
}{.}\hspace{.4pt}1109\discretionary{/}{%
}{/}TVCG\hspace{.1pt}\discretionary{.}{%
}{.}\hspace{.4pt}2020\hspace{.1pt}\discretionary{.}{%
}{.}\hspace{.4pt}3030437}}}


\bibitem{causeinfer2019}
P.~Chen, Y.~Qi, and D.~Hou.
\newblock {CauseInfer}: Automated end-to-end performance diagnosis with hierarchical causality graph in cloud environment.
\newblock {\em IEEE Transactions on Services Computing}, 12(2):214--230, 2019. \href{https://doi.org/10.1109/TSC.2016.2607739}
{doi: {{%
10\hspace{.1pt}\discretionary{.}{%
}{.}\hspace{.4pt}1109\discretionary{/}{%
}{/}TSC\hspace{.1pt}\discretionary{.}{%
}{.}\hspace{.4pt}2016\hspace{.1pt}\discretionary{.}{%
}{.}\hspace{.4pt}2607739}}}


\bibitem{causeinfer2014}
P.~Chen, Y.~Qi, P.~Zheng, and D.~Hou.
\newblock {CauseInfer}: Automatic and distributed performance diagnosis with hierarchical causality graph in large distributed systems.
\newblock In {\em Proceedings of the IEEE Conference on Computer Communications}, pp. 1887--1895, 2014. \href{https://doi.org/10.1109/INFOCOM.2014.6848128}
{doi: {{%
10\hspace{.1pt}\discretionary{.}{%
}{.}\hspace{.4pt}1109\discretionary{/}{%
}{/}INFOCOM\hspace{.1pt}\discretionary{.}{%
}{.}\hspace{.4pt}2014\hspace{.1pt}\discretionary{.}{%
}{.}\hspace{.4pt}6848128}}}


\bibitem{dagre}
Dagre.
\newblock Dagre: Directed graph layout for {JavaScript}.
\newblock \url{https://github.com/dagrejs/dagre}.
\newblock Last accessed: Nov 22, 2023.

\bibitem{GBRCA}
D.~Dandona, M.~Demir, and J.~J. Prevost.
\newblock Graph based root cause analysis in cloud data center.
\newblock In {\em {Proceedings of the IEEE International Conference of System of Systems Engineering}}, pp. 549--554, 2020. \href{https://doi.org/10.1109/SoSE50414.2020.9130526}
{doi: {{%
10\hspace{.1pt}\discretionary{.}{%
}{.}\hspace{.4pt}1109\discretionary{/}{%
}{/}SoSE50414\hspace{.1pt}\discretionary{.}{%
}{.}\hspace{.4pt}2020\hspace{.1pt}\discretionary{.}{%
}{.}\hspace{.4pt}9130526}}}


\bibitem{deng_survey_2023}
Z.~Deng, D.~Weng, S.~Liu, Y.~Tian, M.~Xu, and Y.~Wu.
\newblock A survey of urban visual analytics: {Advances} and future directions.
\newblock {\em Computational Visual Media}, 9(1):3--39, 2023. \href{https://doi.org/10.1007/s41095-022-0275-7}
{doi: {{%
10\hspace{.1pt}\discretionary{.}{%
}{.}\hspace{.4pt}1007\discretionary{/}{%
}{/}s41095\discretionary{%
}{-}{-}022\discretionary{%
}{-}{-}0275\discretionary{%
}{-}{-}7}}}


\bibitem{Fang_2020}
Y.~Fang, H.~Xu, and J.~Jiang.
\newblock A survey of time series data visualization research.
\newblock In {\em Proceedings of the IOP Conference Series: Materials Science and Engineering}, vol. 782, pp. 1--10, 2020. \href{https://doi.org/10.1088/1757-899X/782/2/022013}
{doi: {{%
10\hspace{.1pt}\discretionary{.}{%
}{.}\hspace{.4pt}1088\discretionary{/}{%
}{/}1757\discretionary{%
}{-}{-}899X\discretionary{/}{%
}{/}782\discretionary{/}{%
}{/}2\discretionary{/}{%
}{/}022013}}}


\bibitem{gao2015survey}
Z.~Gao, C.~Cecati, and S.~X. Ding.
\newblock A survey of fault diagnosis and fault-tolerant techniques -- {Part I: Fault} diagnosis with model-based and signal-based approaches.
\newblock {\em IEEE Transactions on Industrial Electronics}, 62(6):3757--3767, 2015. \href{https://doi.org/10.1109/TIE.2015.2417501}
{doi: {{%
10\hspace{.1pt}\discretionary{.}{%
}{.}\hspace{.4pt}1109\discretionary{/}{%
}{/}TIE\hspace{.1pt}\discretionary{.}{%
}{.}\hspace{.4pt}2015\hspace{.1pt}\discretionary{.}{%
}{.}\hspace{.4pt}2417501}}}


\bibitem{aidice}
J.~Gu, C.~Luo, S.~Qin, B.~Qiao, Q.~Lin, H.~Zhang, Z.~Li, Y.~Dang, S.~Cai, W.~Wu, Y.~Zhou, M.~Chintalapati, and D.~Zhang.
\newblock Efficient incident identification from multi-dimensional issue reports via meta-heuristic search.
\newblock In {\em Proceedings of the ACM Joint Meeting on European Software Engineering Conference and Symposium on the Foundations of Software Engineering}, pp. 292--303, 2020. \href{https://doi.org/10.1145/3368089.3409741}
{doi: {{%
10\hspace{.1pt}\discretionary{.}{%
}{.}\hspace{.4pt}1145\discretionary{/}{%
}{/}3368089\hspace{.1pt}\discretionary{.}{%
}{.}\hspace{.4pt}3409741}}}


\bibitem{eventsurvey}
Y.~Guo, S.~Guo, Z.~Jin, S.~Kaul, D.~Gotz, and N.~Cao.
\newblock Survey on visual analysis of event sequence data.
\newblock {\em IEEE Transactions on Visualization and Computer Graphics}, 28(12):5091--5112, 2022. \href{https://doi.org/10.1109/TVCG.2021.3100413}
{doi: {{%
10\hspace{.1pt}\discretionary{.}{%
}{.}\hspace{.4pt}1109\discretionary{/}{%
}{/}TVCG\hspace{.1pt}\discretionary{.}{%
}{.}\hspace{.4pt}2021\hspace{.1pt}\discretionary{.}{%
}{.}\hspace{.4pt}3100413}}}


\bibitem{hagemann2020systematic}
T.~Hagemann and K.~Katsarou.
\newblock A systematic review on anomaly detection for cloud computing environments.
\newblock In {\em Proceedings of the Artificial Intelligence and Cloud Computing Conference}, pp. 83--96, 2020. \href{https://doi.org/10.1145/3442536.3442550}
{doi: {{%
10\hspace{.1pt}\discretionary{.}{%
}{.}\hspace{.4pt}1145\discretionary{/}{%
}{/}3442536\hspace{.1pt}\discretionary{.}{%
}{.}\hspace{.4pt}3442550}}}


\bibitem{hashem2015rise}
I.~A.~T. Hashem, I.~Yaqoob, N.~B. Anuar, S.~Mokhtar, A.~Gani, and S.~{Ullah Khan}.
\newblock The rise of ``big data'' on cloud computing: {R}eview and open research issues.
\newblock {\em {Information Systems}}, 47:98--115, Jan. 2015. \href{https://doi.org/10.1016/j.is.2014.07.006}
{doi: {{%
10\hspace{.1pt}\discretionary{.}{%
}{.}\hspace{.4pt}1016\discretionary{/}{%
}{/}j\hspace{.1pt}\discretionary{.}{%
}{.}\hspace{.4pt}is\hspace{.1pt}\discretionary{.}{%
}{.}\hspace{.4pt}2014\hspace{.1pt}\discretionary{.}{%
}{.}\hspace{.4pt}07\hspace{.1pt}\discretionary{.}{%
}{.}\hspace{.4pt}006}}}


\bibitem{surveyRCAGeneral}
S.~P. Kavulya, K.~Joshi, F.~D. Giandomenico, and P.~Narasimhan.
\newblock Failure diagnosis of complex systems.
\newblock In {\em Resilience Assessment and Evaluation of Computing Systems}, pp. 239--261. 2012. \href{https://doi.org/10.1007/978-3-642-29032-9_12}
{doi: {{%
10\hspace{.1pt}\discretionary{.}{%
}{.}\hspace{.4pt}1007\discretionary{/}{%
}{/}978\discretionary{%
}{-}{-}3\discretionary{%
}{-}{-}642\discretionary{%
}{-}{-}29032\discretionary{%
}{-}{-}9\_12}}}


\bibitem{ViSRE}
P.~Kayongo, J.~Hoffswell, S.~Saini, S.~Garg, E.~Koh, H.~Wang, and T.~Jacobs.
\newblock {ViSRE}: {A} unified visual analysis dashboard for proactive cloud outage management.
\newblock In {\em Proceedings of the Working Conference on Software Visualization}, pp. 5--16, 2022. \href{https://doi.org/10.1109/VISSOFT55257.2022.00010}
{doi: {{%
10\hspace{.1pt}\discretionary{.}{%
}{.}\hspace{.4pt}1109\discretionary{/}{%
}{/}VISSOFT55257\hspace{.1pt}\discretionary{.}{%
}{.}\hspace{.4pt}2022\hspace{.1pt}\discretionary{.}{%
}{.}\hspace{.4pt}00010}}}


\bibitem{compromise}
S.~Kelly.
\newblock Compromise: Modest natural language processing.
\newblock \url{https://github.com/spencermountain/compromise}.
\newblock Last accessed: Nov 22, 2023.

\bibitem{ccbusinesssurvey}
A.~Keshavarzi, A.~T. Haghighat, and M.~Bohlouli.
\newblock Research challenges and prospective business impacts of cloud computing: {A} survey.
\newblock In {\em Proceedings of the {IEEE} International Conference on Intelligent Data Acquisition and Advanced Computing Systems}, pp. 731--736, 2013. \href{https://doi.org/10.1109/IDAACS.2013.6663021}
{doi: {{%
10\hspace{.1pt}\discretionary{.}{%
}{.}\hspace{.4pt}1109\discretionary{/}{%
}{/}IDAACS\hspace{.1pt}\discretionary{.}{%
}{.}\hspace{.4pt}2013\hspace{.1pt}\discretionary{.}{%
}{.}\hspace{.4pt}6663021}}}


\bibitem{monitorrank}
M.~Kim, R.~Sumbaly, and S.~Shah.
\newblock Root cause detection in a service-oriented architecture.
\newblock In {\em Proceedings of the ACM International Conference on Measurement and Modeling of Computer Systems}, pp. 93--104, 2013. \href{https://doi.org/10.1145/2465529.2465753}
{doi: {{%
10\hspace{.1pt}\discretionary{.}{%
}{.}\hspace{.4pt}1145\discretionary{/}{%
}{/}2465529\hspace{.1pt}\discretionary{.}{%
}{.}\hspace{.4pt}2465753}}}


\bibitem{RetainVis}
B.~C. Kwon, M.-J. Choi, J.~T. Kim, E.~Choi, Y.~B. Kim, S.~Kwon, J.~Sun, and J.~Choo.
\newblock {RetainVis}: Visual analytics with interpretable and interactive recurrent neural networks on electronic medical records.
\newblock {\em IEEE Transactions on Visualization and Computer Graphics}, 25(1):299--309, 2019. \href{https://doi.org/10.1109/TVCG.2018.2865027}
{doi: {{%
10\hspace{.1pt}\discretionary{.}{%
}{.}\hspace{.4pt}1109\discretionary{/}{%
}{/}TVCG\hspace{.1pt}\discretionary{.}{%
}{.}\hspace{.4pt}2018\hspace{.1pt}\discretionary{.}{%
}{.}\hspace{.4pt}2865027}}}


\bibitem{lin2018microscope}
J.~Lin, P.~Chen, and Z.~Zheng.
\newblock Microscope: Pinpoint performance issues with causal graphs in micro-service environments.
\newblock In {\em Proceedings of the International Conference on Service-Oriented Computing}, pp. 3--20, 2018. \href{https://doi.org/10.1007/978-3-030-03596-9_1}
{doi: {{%
10\hspace{.1pt}\discretionary{.}{%
}{.}\hspace{.4pt}1007\discretionary{/}{%
}{/}978\discretionary{%
}{-}{-}3\discretionary{%
}{-}{-}030\discretionary{%
}{-}{-}03596\discretionary{%
}{-}{-}9\_1}}}


\bibitem{lin2004viztree}
J.~Lin, E.~Keogh, S.~Lonardi, J.~P. Lankford, and D.~M. Nystrom.
\newblock {VizTree}: A tool for visually mining and monitoring massive time series databases.
\newblock In {\em Proceedings of the International Conference on Very Large Data Bases}, pp. 1269--1272, 2004. \href{https://doi.org/10.5555/1316689.1316811}
{doi: {{%
10\hspace{.1pt}\discretionary{.}{%
}{.}\hspace{.4pt}5555\discretionary{/}{%
}{/}1316689\hspace{.1pt}\discretionary{.}{%
}{.}\hspace{.4pt}1316811}}}


\bibitem{aadrca}
J.~Lin, Q.~Zhang, H.~Bannazadeh, and A.~Leon{-}Garcia.
\newblock Automated anomaly detection and root cause analysis in virtualized cloud infrastructures.
\newblock In {\em {Proceedings of the {IEEE/IFIP} Network Operations and Management Symposium}}, pp. 550--556, 2016. \href{https://doi.org/10.1109/NOMS.2016.7502857}
{doi: {{%
10\hspace{.1pt}\discretionary{.}{%
}{.}\hspace{.4pt}1109\discretionary{/}{%
}{/}NOMS\hspace{.1pt}\discretionary{.}{%
}{.}\hspace{.4pt}2016\hspace{.1pt}\discretionary{.}{%
}{.}\hspace{.4pt}7502857}}}


\bibitem{traceanomaly}
P.~Liu, H.~Xu, Q.~Ouyang, R.~Jiao, Z.~Chen, S.~Zhang, J.~Yang, L.~Mo, J.~Zeng, W.~Xue, and D.~Pei.
\newblock Unsupervised detection of microservice trace anomalies through service-level deep bayesian networks.
\newblock In {\em Proceedings of the IEEE International Symposium on Software Reliability Engineering}, pp. 48--58, 2020. \href{https://doi.org/10.1109/ISSRE5003.2020.00014}
{doi: {{%
10\hspace{.1pt}\discretionary{.}{%
}{.}\hspace{.4pt}1109\discretionary{/}{%
}{/}ISSRE5003\hspace{.1pt}\discretionary{.}{%
}{.}\hspace{.4pt}2020\hspace{.1pt}\discretionary{.}{%
}{.}\hspace{.4pt}00014}}}


\bibitem{ECoalVis}
S.~Liu, D.~Weng, Y.~Tian, Z.~Deng, H.~Xu, X.~Zhu, H.~Yin, X.~Zhan, and Y.~Wu.
\newblock {ECoalVis}: Visual analysis of control strategies in coal-fired power plants.
\newblock {\em IEEE Transactions on Visualization and Computer Graphics}, 29(1):1091--1101, 2023. \href{https://doi.org/10.1109/TVCG.2022.3209430}
{doi: {{%
10\hspace{.1pt}\discretionary{.}{%
}{.}\hspace{.4pt}1109\discretionary{/}{%
}{/}TVCG\hspace{.1pt}\discretionary{.}{%
}{.}\hspace{.4pt}2022\hspace{.1pt}\discretionary{.}{%
}{.}\hspace{.4pt}3209430}}}


\bibitem{corr}
C.~Luo, J.-G. Lou, Q.~Lin, Q.~Fu, R.~Ding, D.~Zhang, and Z.~Wang.
\newblock Correlating events with time series for incident diagnosis.
\newblock In {\em Proceedings of the ACM International Conference on Knowledge Discovery and Data Mining}, pp. 1583--1592, 2014. \href{https://doi.org/10.1145/2623330.2623374}
{doi: {{%
10\hspace{.1pt}\discretionary{.}{%
}{.}\hspace{.4pt}1145\discretionary{/}{%
}{/}2623330\hspace{.1pt}\discretionary{.}{%
}{.}\hspace{.4pt}2623374}}}


\bibitem{kusto}
Microsoft.
\newblock Kusto query language ({KQL}) overview.
\newblock \url{https://learn.microsoft.com/en-us/azure/data-explorer/kusto/query/}.
\newblock Last accessed: Sep 11, 2023.

\bibitem{beline}
C.~Muelder, B.~Zhu, W.~Chen, H.~Zhang, and K.-L. Ma.
\newblock Visual analysis of cloud computing performance using behavioral lines.
\newblock {\em {IEEE Transactions on Visualization and Computer Graphics}}, 22(6):1694--1704, 2016. \href{https://doi.org/10.1109/TVCG.2016.2534558}
{doi: {{%
10\hspace{.1pt}\discretionary{.}{%
}{.}\hspace{.4pt}1109\discretionary{/}{%
}{/}TVCG\hspace{.1pt}\discretionary{.}{%
}{.}\hspace{.4pt}2016\hspace{.1pt}\discretionary{.}{%
}{.}\hspace{.4pt}2534558}}}


\bibitem{riskpipeline}
W.~K. Muhlbauer.
\newblock {Risk: Theory and Application}.
\newblock In {\em Pipeline Risk Management Manual (Third Edition)}, pp. 1--19. 2004. \href{https://doi.org/10.1016/B978-075067579-6/50004-2}
{doi: {{%
10\hspace{.1pt}\discretionary{.}{%
}{.}\hspace{.4pt}1016\discretionary{/}{%
}{/}B978\discretionary{%
}{-}{-}075067579\discretionary{%
}{-}{-}6\discretionary{/}{%
}{/}50004\discretionary{%
}{-}{-}2}}}


\bibitem{FChain}
H.~Nguyen, Z.~Shen, Y.~Tan, and X.~Gu.
\newblock {FChain}: {T}oward black-box online fault localization for cloud systems.
\newblock In {\em Proceedings of the IEEE International Conference on Distributed Computing Systems}, pp. 21--30, 2013. \href{https://doi.org/10.1109/ICDCS.2013.26}
{doi: {{%
10\hspace{.1pt}\discretionary{.}{%
}{.}\hspace{.4pt}1109\discretionary{/}{%
}{/}ICDCS\hspace{.1pt}\discretionary{.}{%
}{.}\hspace{.4pt}2013\hspace{.1pt}\discretionary{.}{%
}{.}\hspace{.4pt}26}}}


\bibitem{PAL}
H.~Nguyen, Y.~Tan, and X.~Gu.
\newblock {PAL: P}ropagation-aware anomaly localization for cloud hosted distributed applications.
\newblock In {\em Proceedings of the ACM Symposium on Operating Systems Principles}, pp. 1--8, 2011. \href{https://doi.org/10.1145/2038633.2038634}
{doi: {{%
10\hspace{.1pt}\discretionary{.}{%
}{.}\hspace{.4pt}1145\discretionary{/}{%
}{/}2038633\hspace{.1pt}\discretionary{.}{%
}{.}\hspace{.4pt}2038634}}}


\bibitem{flask}
Pallets.
\newblock Welcome to {Flask}.
\newblock \url{https://flask.palletsprojects.com}.
\newblock Last accessed: Nov 22, 2023.

\bibitem{MLPDetection}
N.~Pandeeswari and G.~Kumar.
\newblock Anomaly detection system in cloud environment using fuzzy clustering based {ANN}.
\newblock {\em Mobile Networks and Applications}, 21(3):494--505, 2016. \href{https://doi.org/10.1007/s11036-015-0644-x}
{doi: {{%
10\hspace{.1pt}\discretionary{.}{%
}{.}\hspace{.4pt}1007\discretionary{/}{%
}{/}s11036\discretionary{%
}{-}{-}015\discretionary{%
}{-}{-}0644\discretionary{%
}{-}{-}x}}}


\bibitem{IncidentAnomaly}
C.~Raj, L.~Khular, and G.~Raj.
\newblock Clustering based incident handling for anomaly detection in cloud infrastructures.
\newblock In {\em Proceedings of the International Conference on Cloud Computing, Data Science \& Engineering (Confluence)}, pp. 611--616, 2020. \href{https://doi.org/10.1109/Confluence47617.2020.9058314}
{doi: {{%
10\hspace{.1pt}\discretionary{.}{%
}{.}\hspace{.4pt}1109\discretionary{/}{%
}{/}Confluence47617\hspace{.1pt}\discretionary{.}{%
}{.}\hspace{.4pt}2020\hspace{.1pt}\discretionary{.}{%
}{.}\hspace{.4pt}9058314}}}


\bibitem{bocp}
Redpoll.
\newblock {A Bayesian change point library}.
\newblock \url{https://pypi.org/project/changepoint/}.
\newblock Last accessed: Nov 22, 2023.

\bibitem{rosenthal2013visruption}
P.~Rosenthal, L.~Pfeiffer, N.~H. M{\"u}ller, and P.~Ohler.
\newblock {VisRuption}: Intuitive and efficient visualization of temporal airline disruption data.
\newblock {\em Computer Graphics Forum}, 32:81--90, 2013. \href{https://doi.org/10.1111/cgf.12095}
{doi: {{%
10\hspace{.1pt}\discretionary{.}{%
}{.}\hspace{.4pt}1111\discretionary{/}{%
}{/}cgf\hspace{.1pt}\discretionary{.}{%
}{.}\hspace{.4pt}12095}}}


\bibitem{Traveler}
S.~A. Sakin, A.~Bigelow, R.~Tohid, C.~Scully-Allison, C.~Scheidegger, S.~R. Brandt, C.~Taylor, K.~A. Huck, H.~Kaiser, and K.~E. Isaacs.
\newblock Traveler: Navigating task parallel traces for performance analysis.
\newblock {\em IEEE Transactions on Visualization and Computer Graphics}, 29(1):788--797, 2023. \href{https://doi.org/10.1109/TVCG.2022.3209375}
{doi: {{%
10\hspace{.1pt}\discretionary{.}{%
}{.}\hspace{.4pt}1109\discretionary{/}{%
}{/}TVCG\hspace{.1pt}\discretionary{.}{%
}{.}\hspace{.4pt}2022\hspace{.1pt}\discretionary{.}{%
}{.}\hspace{.4pt}3209375}}}


\bibitem{DLA}
A.~Samir and C.~Pahl.
\newblock {DLA}: Detecting and localizing anomalies in containerized microservice architectures using markov models.
\newblock In {\em Proceedings of the International Conference on Future Internet of Things and Cloud}, pp. 205--213, 2019. \href{https://doi.org/10.1109/FiCloud.2019.00036}
{doi: {{%
10\hspace{.1pt}\discretionary{.}{%
}{.}\hspace{.4pt}1109\discretionary{/}{%
}{/}FiCloud\hspace{.1pt}\discretionary{.}{%
}{.}\hspace{.4pt}2019\hspace{.1pt}\discretionary{.}{%
}{.}\hspace{.4pt}00036}}}


\bibitem{designmethod}
M.~Sedlmair, M.~Meyer, and T.~Munzner.
\newblock Design study methodology: Reflections from the trenches and the stacks.
\newblock {\em IEEE Transactions on Visualization and Computer Graphics}, 18(12):2431--2440, 2012. \href{https://doi.org/10.1109/TVCG.2012.213}
{doi: {{%
10\hspace{.1pt}\discretionary{.}{%
}{.}\hspace{.4pt}1109\discretionary{/}{%
}{/}TVCG\hspace{.1pt}\discretionary{.}{%
}{.}\hspace{.4pt}2012\hspace{.1pt}\discretionary{.}{%
}{.}\hspace{.4pt}213}}}


\bibitem{shanplatform}
H.~Shan, Y.~Chen, H.~Liu, Y.~Zhang, X.~Xiao, X.~He, M.~Li, and W.~Ding.
\newblock $\epsilon$-{D}iagnosis: Unsupervised and real-time diagnosis of small-window long-tail latency in large-scale microservice platforms.
\newblock In {\em Proceedings of the World Wide Web Conference}, pp. 3215--3222, 2019. \href{https://doi.org/10.1145/3308558.3313653}
{doi: {{%
10\hspace{.1pt}\discretionary{.}{%
}{.}\hspace{.4pt}1145\discretionary{/}{%
}{/}3308558\hspace{.1pt}\discretionary{.}{%
}{.}\hspace{.4pt}3313653}}}


\bibitem{Hardware}
Shilpika, T.~Fujiwara, N.~Sakamoto, J.~Nonaka, and K.-L. Ma.
\newblock A visual analytics approach for hardware system monitoring with streaming functional data analysis.
\newblock {\em IEEE Transactions on Visualization and Computer Graphics}, 28(6):2338--2349, 2022. \href{https://doi.org/10.1109/TVCG.2022.3165348}
{doi: {{%
10\hspace{.1pt}\discretionary{.}{%
}{.}\hspace{.4pt}1109\discretionary{/}{%
}{/}TVCG\hspace{.1pt}\discretionary{.}{%
}{.}\hspace{.4pt}2022\hspace{.1pt}\discretionary{.}{%
}{.}\hspace{.4pt}3165348}}}


\bibitem{networksurvey}
H.~Shiravi, A.~Shiravi, and A.~A. Ghorbani.
\newblock A survey of visualization systems for network security.
\newblock {\em IEEE Transactions on Visualization and Computer Graphics}, 18(8):1313--1329, 2012. \href{https://doi.org/10.1109/TVCG.2011.144}
{doi: {{%
10\hspace{.1pt}\discretionary{.}{%
}{.}\hspace{.4pt}1109\discretionary{/}{%
}{/}TVCG\hspace{.1pt}\discretionary{.}{%
}{.}\hspace{.4pt}2011\hspace{.1pt}\discretionary{.}{%
}{.}\hspace{.4pt}144}}}


\bibitem{cloudRCAsurvey}
J.~Soldani and A.~Brogi.
\newblock Anomaly detection and failure root cause analysis in (micro) service-based cloud applications: A survey.
\newblock {\em ACM Computing Surveys}, 55(3):59:1--59:39, Feb 2022. \href{https://doi.org/10.1145/3501297}
{doi: {{%
10\hspace{.1pt}\discretionary{.}{%
}{.}\hspace{.4pt}1145\discretionary{/}{%
}{/}3501297}}}


\bibitem{surveyRCAInfer}
M.~Sol{\'{e}}, V.~Munt{\'{e}}s{-}Mulero, A.~I. Rana, and G.~Estrada.
\newblock Survey on models and techniques for root-cause analysis.
\newblock {\em CoRR}, abs/1701.08546, 2017. \href{https://doi.org/10.48550/arXiv.1701.08546}
{doi: {{%
10\hspace{.1pt}\discretionary{.}{%
}{.}\hspace{.4pt}48550\discretionary{/}{%
}{/}arXiv\hspace{.1pt}\discretionary{.}{%
}{.}\hspace{.4pt}1701\hspace{.1pt}\discretionary{.}{%
}{.}\hspace{.4pt}08546}}}


\bibitem{planningvis}
D.~Sun, R.~Huang, Y.~Chen, Y.~Wang, J.~Zeng, M.~Yuan, T.-C. Pong, and H.~Qu.
\newblock {PlanningVis}: A visual analytics approach to production planning in smart factories.
\newblock {\em IEEE Transactions on Visualization and Computer Graphics}, 26(1):579--589, 2020. \href{https://doi.org/10.1109/TVCG.2019.2934275}
{doi: {{%
10\hspace{.1pt}\discretionary{.}{%
}{.}\hspace{.4pt}1109\discretionary{/}{%
}{/}TVCG\hspace{.1pt}\discretionary{.}{%
}{.}\hspace{.4pt}2019\hspace{.1pt}\discretionary{.}{%
}{.}\hspace{.4pt}2934275}}}


\bibitem{Sieve}
J.~Thalheim, A.~Rodrigues, I.~E. Akkus, P.~Bhatotia, R.~Chen, B.~Viswanath, L.~Jiao, and C.~Fetzer.
\newblock Sieve: Actionable insights from monitored metrics in distributed systems.
\newblock In {\em Proceedings of the ACM/IFIP/USENIX Middleware Conference}, pp. 14--27, 2017. \href{https://doi.org/10.1145/3135974.3135977}
{doi: {{%
10\hspace{.1pt}\discretionary{.}{%
}{.}\hspace{.4pt}1145\discretionary{/}{%
}{/}3135974\hspace{.1pt}\discretionary{.}{%
}{.}\hspace{.4pt}3135977}}}


\bibitem{tominski2004axes}
C.~Tominski, J.~Abello, and H.~Schumann.
\newblock Axes-based visualizations with radial layouts.
\newblock In {\em Proceedings of the ACM Symposium on Applied Computing}, pp. 1242--1247, 2004. \href{https://doi.org/10.1145/967900.968153}
{doi: {{%
10\hspace{.1pt}\discretionary{.}{%
}{.}\hspace{.4pt}1145\discretionary{/}{%
}{/}967900\hspace{.1pt}\discretionary{.}{%
}{.}\hspace{.4pt}968153}}}


\bibitem{GRANO}
H.~Wang, P.~Nguyen, J.~Li, S.~Kopru, G.~Zhang, S.~Katariya, and S.~Ben-Romdhane.
\newblock {GRANO}: Interactive graph-based root cause analysis for cloud-native distributed data platform.
\newblock {\em Proceedings of the VLDB Endowment}, 12(12):1942--1945, 2019. \href{https://doi.org/10.14778/3352063.3352105}
{doi: {{%
10\hspace{.1pt}\discretionary{.}{%
}{.}\hspace{.4pt}14778\discretionary{/}{%
}{/}3352063\hspace{.1pt}\discretionary{.}{%
}{.}\hspace{.4pt}3352105}}}


\bibitem{wangservice}
L.~Wang, N.~Zhao, J.~Chen, P.~Li, W.~Zhang, and K.~Sui.
\newblock Root-cause metric location for microservice systems via log anomaly detection.
\newblock In {\em Proceedings of the IEEE International Conference on Web Services}, pp. 142--150, 2020. \href{https://doi.org/10.1109/ICWS49710.2020.00026}
{doi: {{%
10\hspace{.1pt}\discretionary{.}{%
}{.}\hspace{.4pt}1109\discretionary{/}{%
}{/}ICWS49710\hspace{.1pt}\discretionary{.}{%
}{.}\hspace{.4pt}2020\hspace{.1pt}\discretionary{.}{%
}{.}\hspace{.4pt}00026}}}


\bibitem{citenetwork}
T.~Wang, Z.~Li, and J.~Zhang.
\newblock Egocentric visual analysis of dynamic citation network.
\newblock {\em Journal of Visualization}, 25(6):1343--1360, 2022. \href{https://doi.org/10.1007/s12650-022-00862-7}
{doi: {{%
10\hspace{.1pt}\discretionary{.}{%
}{.}\hspace{.4pt}1007\discretionary{/}{%
}{/}s12650\discretionary{%
}{-}{-}022\discretionary{%
}{-}{-}00862\discretionary{%
}{-}{-}7}}}


\bibitem{downtimefor}
S.~Wastie.
\newblock The real cost of downtime, the real potential of {DevOps}.
\newblock AppDynamics, Jul 2018.
\newblock Available: \url{https://www.appdynamics.com/blog/engineering/idc-devops-cost-downtime/} (Last accessed: Nov 22, 2024).

\bibitem{amazonincidents}
S.~Wolfe.
\newblock {Amazon's one hour of downtime on Prime Day may have cost it up to \$100 million in lost sales}.
\newblock Business Insider, Jul 2018.
\newblock Available: \url{https://www.businessinsider.com/amazon-prime-day-website-issues-cost-it-millions-in-lost-sales-2018-7} (Last accessed: Mar 30, 2024).

\bibitem{surveyRCAsoftware}
W.~E. Wong, R.~Gao, Y.~Li, R.~Abreu, and F.~Wotawa.
\newblock A survey on software fault localization.
\newblock {\em IEEE Transactions on Software Engineering}, 42(8):707--740, 2016. \href{https://doi.org/10.1109/TSE.2016.2521368}
{doi: {{%
10\hspace{.1pt}\discretionary{.}{%
}{.}\hspace{.4pt}1109\discretionary{/}{%
}{/}TSE\hspace{.1pt}\discretionary{.}{%
}{.}\hspace{.4pt}2016\hspace{.1pt}\discretionary{.}{%
}{.}\hspace{.4pt}2521368}}}


\bibitem{wongsuphasawat2011outflow}
K.~Wongsuphasawat and D.~Gotz.
\newblock Outflow: Visualizing patient flow by symptoms and outcome.
\newblock In {\em Proceedings of the IEEE VisWeek Workshop on Visual Analytics in Healthcare}, pp. 25--28, 2011.

\bibitem{MicroRCA}
L.~Wu, J.~Tordsson, E.~Elmroth, and O.~Kao.
\newblock Micro{RCA}: Root cause localization of performance issues in microservices.
\newblock In {\em Proceedings of the IEEE/IFIP Network Operations and Management Symposium}, pp. 1--9, 2020. \href{https://doi.org/10.1109/NOMS47738.2020.9110353}
{doi: {{%
10\hspace{.1pt}\discretionary{.}{%
}{.}\hspace{.4pt}1109\discretionary{/}{%
}{/}NOMS47738\hspace{.1pt}\discretionary{.}{%
}{.}\hspace{.4pt}2020\hspace{.1pt}\discretionary{.}{%
}{.}\hspace{.4pt}9110353}}}


\bibitem{industrysurvey}
W.~Wu, Y.~Zheng, K.~Chen, X.~Wang, and N.~Cao.
\newblock A visual analytics approach for equipment condition monitoring in smart factories of process industry.
\newblock In {\em Proceedings of the IEEE Pacific Visualization Symposium}, pp. 140--149, 2018. \href{https://doi.org/10.1109/PacificVis.2018.00026}
{doi: {{%
10\hspace{.1pt}\discretionary{.}{%
}{.}\hspace{.4pt}1109\discretionary{/}{%
}{/}PacificVis\hspace{.1pt}\discretionary{.}{%
}{.}\hspace{.4pt}2018\hspace{.1pt}\discretionary{.}{%
}{.}\hspace{.4pt}00026}}}


\bibitem{clouddet}
K.~Xu, Y.~Wang, L.~Yang, Y.~Wang, B.~Qiao, S.~Qin, Y.~Xu, H.~Zhang, and H.~Qu.
\newblock Cloud{D}et: {I}nteractive visual analysis of anomalous performances in cloud computing systems.
\newblock {\em {IEEE Transactions on Visualization and Computer Graphics}}, 26(1):1107--1117, Jan. 2020. \href{https://doi.org/10.1109/TVCG.2019.2934613}
{doi: {{%
10\hspace{.1pt}\discretionary{.}{%
}{.}\hspace{.4pt}1109\discretionary{/}{%
}{/}TVCG\hspace{.1pt}\discretionary{.}{%
}{.}\hspace{.4pt}2019\hspace{.1pt}\discretionary{.}{%
}{.}\hspace{.4pt}2934613}}}


\bibitem{ViDX}
P.~Xu, H.~Mei, L.~Ren, and W.~Chen.
\newblock {ViDX}: Visual diagnostics of assembly line performance in smart factories.
\newblock {\em IEEE Transactions on Visualization and Computer Graphics}, 23(1):291--300, 2017. \href{https://doi.org/10.1109/TVCG.2016.2598664}
{doi: {{%
10\hspace{.1pt}\discretionary{.}{%
}{.}\hspace{.4pt}1109\discretionary{/}{%
}{/}TVCG\hspace{.1pt}\discretionary{.}{%
}{.}\hspace{.4pt}2016\hspace{.1pt}\discretionary{.}{%
}{.}\hspace{.4pt}2598664}}}


\bibitem{vue}
E.~You.
\newblock {Vue.js: The Progressive JavaScript Framework}.
\newblock \url{https://vuejs.org/}.
\newblock Last accessed: Nov 22, 2023.

\bibitem{cloudrca}
Y.~Zhang, Z.~Guan, H.~Qian, L.~Xu, H.~Liu, Q.~Wen, L.~Sun, J.~Jiang, L.~Fan, and M.~Ke.
\newblock Cloud{RCA}: {A} root cause analysis framework for cloud computing platforms.
\newblock In {\em {Proceedings of the ACM International Conference on Information and Knowledge Management}}, pp. 4373--4382, 2021. \href{https://doi.org/10.1145/3459637.3481903}
{doi: {{%
10\hspace{.1pt}\discretionary{.}{%
}{.}\hspace{.4pt}1145\discretionary{/}{%
}{/}3459637\hspace{.1pt}\discretionary{.}{%
}{.}\hspace{.4pt}3481903}}}


\bibitem{manu2019survey}
F.~Zhou, X.~Lin, C.~Liu, Y.~Zhao, P.~Xu, L.~Ren, T.~Xue, and L.~Ren.
\newblock A survey of visualization for smart manufacturing.
\newblock {\em Journal of Visualization}, 22:419--435, 2019. \href{https://doi.org/10.1007/s12650-018-0530-2}
{doi: {{%
10\hspace{.1pt}\discretionary{.}{%
}{.}\hspace{.4pt}1007\discretionary{/}{%
}{/}s12650\discretionary{%
}{-}{-}018\discretionary{%
}{-}{-}0530\discretionary{%
}{-}{-}2}}}


\end{thebibliography}
\end{document}